\documentclass[pra, aps, twocolumn, superscriptaddress, nofootinbib]{revtex4-2}
\usepackage[a4paper, left=2cm, right=2cm, top=2cm, bottom=2cm]{geometry}
\usepackage{float} % For precise placement of tables

\usepackage[english]{babel}
\usepackage[utf8]{inputenc}
\usepackage[T1]{fontenc}
\usepackage{amsmath}
\usepackage{graphicx}
\usepackage{mathtools}
\usepackage[colorinlistoftodos]{todonotes}
\usepackage{xcolor}
\definecolor{prlink}{RGB}{46,48,146}

\usepackage{hyperref}
\hypersetup{colorlinks=true, allcolors=prlink}
\usepackage[nameinlink]{cleveref}

\usepackage{bbm,microtype,mathrsfs,amsmath,amssymb,color,amsthm,graphicx,bm}
\usepackage{tikz-cd}
\usepackage{braket}
\usepackage{thm-restate}

\usepackage{amsthm}
\newtheorem{thm}{Theorem}[section]
\newtheorem{cor}{Corollary}[section]
\newtheorem{lem}{Lemma}[section]

\newtheorem{fact}{Fact}[section]
\newtheorem{prop}{Proposition}[section]

\newtheorem{rmk}{Remark}[thm]
\newtheorem{defn}{Definition}[section]
\usepackage{comment}
\usepackage{float}
\newcommand{\BE }{\mathbb{E}}

\newcommand{\CO}{\mathcal{O}}

\newcommand{\BR}{\mathbb{R}}

\newcommand{\vA}{\bm{A}}

\newcommand{\vH}{\bm{H}}

\newcommand{\vI}{\mathbf{I}}

\newcommand{\vO}{\bm{O}}
\newcommand{\vP}{\bm{P}}

\newcommand{\vX}{\bm{X}}

\newcommand{\vY}{\bm{Y}}
\newcommand{\vZ}{\bm{Z}}
\newcommand{\vsigma}{\bm{\sigma}}
\newcommand{\vrho}{\bm{\rho}}

\renewcommand{\L}{\left}
\newcommand{\R}{\right}

\newcommand*{\tr}{\operatorname{Tr}}
\newcommand*{\btr}{\operatorname{\overline{Tr}}}

\newcommand{\vertiiismall}[1]{{\vert\kern-0.25ex\vert\kern-0.25ex\vert #1 
    \vert\kern-0.25ex\vert\kern-0.25ex\vert}}

\newcommand{\vertiii}[1]{{\left\vert\kern-0.25ex\left\vert\kern-0.25ex\left\vert #1 
    \right\vert\kern-0.25ex\right\vert\kern-0.25ex\right\vert}}
\newcommand{\vertiiiNoLR}[1]{{\bigg\vert\kern-0.25ex\bigg\vert\kern-0.25ex\bigg\vert #1 
    \bigg\vert\kern-0.25ex\bigg\vert\kern-0.25ex\bigg\vert}}
\newcommand{\lvertiii}{\bigg\vert\kern-0.25ex\bigg\vert\kern-0.25ex\bigg\vert }
\newcommand{\rvertiii}{\bigg\vert\kern-0.25ex\bigg\vert\kern-0.25ex\bigg\vert }
\newcommand{\norm}[1]{\Vert {#1} \Vert}

\newcommand{\labs}[1]{\left\vert {#1} \right\vert}

\newcommand{\e}{\mathrm{e}}

\newcommand{\Expect}{\operatorname{\mathbb{E}}}

\DeclareMathOperator{\Tr}{Tr}

\DeclareMathOperator{\chrom}{chrom}
\DeclareMathOperator{\syk}{SYK}

\begin{document}
\title{Strongly interacting fermions are non-trivial yet non-glassy}

\author{Eric R.\ Anschuetz}
\email{eans@caltech.edu}
\affiliation{Institute for Quantum Information and Matter, Caltech, Pasadena, CA, USA}
\affiliation{Walter Burke Institute for Theoretical Physics, Caltech, Pasadena, CA, USA}
\author{Chi-Fang Chen}
\email{achifchen@gmail.com}
\affiliation{Institute for Quantum Information and Matter, Caltech, Pasadena, CA, USA}
\affiliation{University of California, Berkeley, CA, USA}
\author{Bobak T.\ Kiani}
\email{bkiani@seas.harvard.edu}
\affiliation{John A.\ Paulson School of Engineering and Applied Sciences, Harvard, Cambridge, MA, USA}
\author{Robbie King}
\email{wking@caltech.edu}
\affiliation{Institute for Quantum Information and Matter, Caltech, Pasadena, CA, USA}
\affiliation{Computational and Mathematical Sciences, Caltech, Pasadena, CA, USA}

\begin{abstract}
% Shorter abstract
% Random spin systems feature glassy physics and computational hardness. We study random all-to-all fermionic systems (Sachdev-Ye-Kitaev model) and prove that (I) its low-energy states have polynomial circuit complexity, yet (II) annealed and quenched free energies agree to low temperatures, ruling out a glassy phase transition in this sense. Both results are rooted in the non-commutativity of fermionic terms quantified by the commutation index. Our results suggest that strongly interacting fermions at low temperatures, unlike spins, belong in a classically nontrivial yet quantumly easy phase.

% Longer abstract
Random spin systems at low temperatures are glassy and feature computational hardness in finding low-energy states. We study the random all-to-all interacting fermionic Sachdev--Ye--Kitaev (SYK) model and prove that, in contrast, (I) the low-energy states have polynomial circuit depth, yet (II) the annealed and quenched free energies agree to inverse-polynomially low temperatures, ruling out a glassy phase transition in this sense. These results are derived by showing that fermionic and spin systems significantly differ in their \emph{commutation index}, which quantifies the non-commutativity of Hamiltonian terms. Our results suggest that low-temperature strongly interacting fermions, unlike spins, belong in a classically nontrivial yet quantumly easy phase.
\end{abstract}
\maketitle

\section{Introduction}

Simulating ground and thermal state properties of quantum systems is a key application of future quantum computers \cite{feynman1982SimQPhysWithComputers,lloyd1996universal,mcardle2020quantum,THC_google,2021_Microsoft_catalysis,babbush2018low,chamberland2020building}.
Nevertheless, the search for particular, favorable instances that are quantumly easy and classically hard is not clear-cut~\cite{Isthere_22_lee}. A challenge is that current quantum computers are limited in quality and size, requiring the community to rely on theoretical arguments to give computational separations. However, the ground states for standard few-body quantum spin models can be QMA-hard (as classical spin models are NP-hard) in the worst case~\cite{kitaev2002classical,aharonov2009power,gottesman2009quantum}; in the average case, random classical and quantum spin models exhibit glassy physics where computational hardness may arise \cite{gamarnik2021overlap,Swingle20_SK_SYK}. To give an efficient quantum algorithm for low-temperature states, one must carefully avoid these instances.

Most chemical and condensed matter systems involve \textit{fermionic} degrees of freedom, not only spins. Of particular importance in quantum chemistry is the \emph{strongly interacting} regime, where Gaussian states do not give good approximations to the ground state and the Hartree--Fock method fails \cite{szabo2012modern}. This has been proposed as a promising regime in which to apply quantum computers to achieve quantum advantage \cite{mcardle2020quantum}. The Sachdev--Ye--Kitaev (SYK) Hamiltonian provides a natural model for strongly interacting fermions \cite{sachdev1993gapless,kitaev2015hidden,kitaev2015simple}. As a counterpart to random spins, it is a random Hamiltonian consisting of all-to-all $q$-body Majorana fermions:
\begin{align}\label{eq:syk_model}
    \vH^{\text{SYK}}_{q} := i^{q/2}{n \choose q}^{-1/2} \sum_{j_1<\dots<j_q} g_{j_1 \dots j_q} \gamma_{j_1} \dots \gamma_{j_q}
\end{align}
where $q$ is assumed to be even, $\gamma_i\gamma_j+\gamma_j\gamma_i = 2\delta_{ij}$, and the $g_{j_1 \dots j_q}$ are i.i.d.~standard Gaussian random variables.

While the 4-body fermionic ground state problem can be just as hard as spin models in the worst case (NP-hard) \cite{liu2007quantum}, average-case fermionic systems appear to have qualitatively different physics and perhaps computational complexity than spin systems \cite{hastings2022optimizing,maldacena1604comments,Swingle20_SK_SYK}. Extensive heuristic calculations (such as large-N expansions) together with numerical evidence indicate that the SYK model resembles a thermalizing chaotic system, not a frozen spin-glass as occurs with few-body quantum spin systems~\cite{Swingle20_SK_SYK,facoetti2019classical}. However, rigorous proofs that go beyond the physical arguments have been very limited~\cite{hastings2022optimizing,feng2019spectrum}. 

In this Letter, we study the strongly interacting SYK model and give quantitative evidence that random, all-to-all connected fermionic systems have a \textit{classically non-trivial} yet \textit{non-glassy} thermal state at constant temperatures. In contrast, these two properties are false for disordered spin systems~\cite{baldwin2020quenched,bravyi2019approximation}. Remarkably, the proofs of both main results rely on the same quantity, the \emph{commutation index}~\cite{king2024triply}. To bound the commutation index of fermionic operators, we analyze the \emph{Lovasz theta-function}~\cite{knuth1993sandwich} of a certain graph encoding the fermionic commutation relations.

This quantity pinpoints a crucial and often overlooked distinction between fermionic and spin Hamiltonians: low-degree fermionic monomials have a very different commutation structure than low-weight Pauli operators. The commutation index captures this difference, quantifying the fundamental distinction in the physics of local spin systems and local fermionic systems. This disparity, we argue, is the origin of a potential quantum advantage in simulating strongly interacting fermionic systems.

More precisely, we first show that all low-energy states (including constant-temperature thermal states) of the SYK model have high circuit complexity (`classically non-trivial'):

\begin{thm}[Low energy states are classically nontrivial]\label{thm:intro_NLTS}
    Consider the degree-$q$ SYK model $\vH^{\syk}_q$. With high probability, the maximum energy is $\lambda_{\max}(\vH^{\syk}_q) \geq \Omega_q(\sqrt{n})$, yet any state $\bm{\rho}$ such that
    \begin{equation}
    \Tr\left(\bm{\rho}\vH^{\syk}_q\right) \geq t \sqrt{n}
    \end{equation}
    has circuit complexity
    \begin{equation} \label{eq:circuit_complexity}
    \tilde{\Omega}_q(n^{(q/2)+1} t^2).
    \end{equation}
    The $\Omega_q$ notations assume a fixed $q$ and growing $n$.
\end{thm}
That is, low-energy states of the SYK model are highly entangled and require many parameters to describe; simple classical ansatzes, such as Gaussian states, must fail. In comparison, local quantum spin systems are known to have efficiently computable product state approximations to the ground state~\cite{bravyi2019approximation} and thus, in this sense, have `trivial' states that achieve a constant-factor approximation of the ground state energy.

Second, we show that the quenched free energy of the SYK model agrees with the annealed free energy even at very low temperatures (`non-glassy'), formalizing and strengthening previous results of this nature \cite{gurau2017quenched,Swingle20_SK_SYK,facoetti2019classical,georges2000mean}.\footnote{In particular, Ref.~\cite{Swingle20_SK_SYK} showed that the SYK model is \textit{consistent} with an annealed approximation, and here we prove that the annealed approximation \textit{holds}.} Here, the free energy is normalized such that $\beta = \CO(1)$ corresponds to constant physical temperature. 

\begin{thm}[Annealed at low temperatures]\label{thm:annealed_low_temps}
    Consider the partition function of the degree-$q$ SYK model $Z_\beta := \Tr{\exp(-\beta \sqrt{n} \vH^{\syk}_q)}$. Then, we have:
    \begin{align}
    \frac{\BE \ln{Z_\beta}}{n} \le \frac{\ln \BE Z_\beta}{n} \le \frac{\BE \ln{Z_\beta}}{n} + \mathcal{O}_q(\beta^2 n^{-q/2}).
    \end{align}
    The $\CO_q$ notations assume a fixed $q$ and growing $n$.
\end{thm}
The quantitative agreement of the two free energies at (inverse-polynomially) low temperatures strikes a stark contrast with disordered spin systems: the SYK model does not experience a `glass' phase transition in the sense of quenched-vs.-annealed free energy. For classical spin Hamiltonians, it is known that the annealed free energy $n^{-1}\BE \ln{Z_\beta}$ fails to agree with the quenched free energy $n^{-1}\ln{\BE Z_\beta}$ at constant temperatures where the Hamiltonian is in its glassy phase and algorithmic hardness arises (see \Cref{app:spin_glass_background}); disordered quantum spin systems undergo a similar transition at constant temperature~\cite{baldwin2020quenched}. The lack of a glass transition for the SYK model suggests that there may be no algorithmic obstructions to preparing low-temperature states of the model on a quantum computer, but we do not prove this claim. We leave finding such an efficient quantum algorithm for future work.

Finally, we study the ground state energy of the SYK model as a function of the locality $q$, as an extension of Theorem~\ref{thm:intro_NLTS} with potentially large $q$.
\begin{thm}[Lower bounding the norm with $q$-dependence]
    For every $q,n$, it holds that the maximum energy of the degree-$q$ SYK model is $\BE\lambda_{\text{max}} \geq \Omega(\sqrt{n}/q)$.
\end{thm}
We show a similar $\Omega(\sqrt{n}/q)$ scaling for $q$-body quantum spin glasses. To our knowledge this is the first \emph{lower bound} on the maximum eigenvalue which scales as $\sqrt{n}/q$; for both models only the scaling at constant $q$ was previously known~\cite{feng2019spectrum,hastings2022optimizing,herasymenko2023optimizing,anschuetz2024bounds}. Our lower bound technique relies on a measure of anticommutation which we call the \emph{commutation degree}, which counts the maximum number of operators that anti-commute with any given operator in the Hamiltonian. Interestingly, the commutation degree does not distinguish between local spin operators and local fermionic operators.

\emph{Background and related work.} \quad
The SYK model is a canonical instance of a chaotic Hamiltonian \cite{sachdev1993gapless,kitaev2015hidden,kitaev2015simple} with related models studied as far back as \cite{french1970validity,bohigas1971two}. For even $q = o(\sqrt{n})$, the SYK model has a Gaussian spectrum \cite{feng2019spectrum} and heuristics from physics indicate that the expected maximum energy of the SYK model scales as $\frac{\sqrt{2n}}{q}$ for even $q$ \cite{garcia2018exact,garcia2016spectral,hastings2022optimizing}. However, the only rigorous result we are aware of with explicit constants is an upper bound of $\sqrt{\log(2) n}$ \cite{feng2019spectrum}. Though Gaussian state approximation algorithms exist for fermionic systems \cite{bravyi2019approximation,herasymenko2023optimizing}, it is known that for the SYK model with $q \geq 4$, Gaussian states cannot achieve constant factor approximations to the maximum energy \cite{haldar2021variational}. Separate from the SYK model, so-called no low-energy trivial states (NLTS) theorems rule out constant factor approximations to ground energies with low depth circuits in worst-case settings \cite{anshu2023nlts,herasymenko2023fermionic}. For random nonlocal Hamiltonians, \cite{chen2023sparse} shows a circuit lower bound for sparse, sampled Pauli models using a similar technique as in the proof of \autoref{thm:intro_NLTS}, where the commutation index is much more straightforward to calculate. 

The commutation index has connections to other areas of quantum information theory and Hamiltonian complexity. In \cite{de2023uncertainty,xu2023bounding}, the commutation index (there termed the \emph{generalized radius}) is used to study generalized Heisenberg uncertainty relations. Related to our work, \cite{hastings2022optimizing} use the commutation index to analyze the performance of sum-of-squares relaxations of the SYK model and prove the $q=4$ instance of \autoref{thm:main_informal}, giving as well an algorithm verifying $\Omega(\sqrt{n})$ energy for $q=4$. 
\cite{anschuetz2024bounds} demonstrates that product states maximize the energy variance for random quantum spin Hamiltonians.
Finally, the commutation index appears in quantum learning theory, where it provides a sample-complexity lower bound on how many copies of the state are required to learn the expectation values of a set of operators via shadow tomography \cite{chen2022exponential,king2024triply}.

\section{Commutation structure of local operators} \label{sec:commutation}

The \emph{commutation index} $\Delta(\mathcal{S})$ of a given set of operators $\mathcal{S}=\{\vA_1,\dots,\vA_m\}$ is defined to be~\cite{king2024triply}:
\begin{equation}
    \Delta(\mathcal{S}) := \sup_{|\psi\rangle} \frac{1}{m} \sum_{i=1}^m \langle\psi|\vA_i|\psi\rangle^2.
\end{equation}
When all $\left\lVert\vA_i\right\rVert\leq 1$ the commutation index takes values $0 < \Delta(\mathcal{S}) \leq 1$. Roughly, a more `commuting' set of observables $\mathcal{S}$ gives a larger value of $\Delta\left(\mathcal{S}\right)$. For example, if the operators in $\mathcal{S}$ are all mutually commuting Pauli operators, choosing $|\psi\rangle$ to be a simultaneous eigenstate gives $\Delta(\mathcal{S}) = 1$.

The commutation index has strong implications for the physics of the model $\vH=m^{-1/2}\sum_{i=1}^m g_i \vA_i$ with Gaussian coefficients $g_i$. Crucially, it controls the sensitivity of many physical properties when varying the couplings of the model. For instance, the norm of the energy gradient of a given state with respect to the disorder is bounded by:
\begin{equation}
    \big|\big|\bm{\nabla}_{\bm{g}} \langle\phi|\vH|\phi\rangle\big|\big|_2^2 = \frac{1}{m} \sum_{i=1}^m \langle\phi|\vA_i|\phi\rangle^2 \leq \Delta(\mathcal{S}).
\end{equation}

Our key observation is that the commutation index of the set $\mathcal{S}^n_q$ of ${n \choose q}$ degree-$q$ Majorana operators is very small:
\begin{thm} \label{thm:main_informal}
Let $\mathcal{S}^n_q$ be the set of degree-$q$ Majorana operators on $n$ fermionic modes. Then for any constant, even $q$:
\begin{equation} \label{eq:main_informal}
\Delta(\mathcal{S}^n_q) = \Theta_q(n^{-q/2}).
\end{equation}
\end{thm}

The decay with system size $n$ is unique to the fermionic setting---for local Pauli operators, $\Delta(\mathcal{S})$ is constant with respect to $n$ (see \autoref{tab:comm_index_results}). This behaviour was first conjectured in \cite{hastings2022optimizing} to our knowledge, and we establish the conjecture---including the setting when $q$ scales with $n$---in \Cref{app:majorana}.

\begin{table}
\centering
\begin{tabular}{lc}
\hline
\textbf{Set $\mathcal{S}$} & \textbf{Commutation index $\Delta(\mathcal{S})$} \\
\hline
Commuting & $1$ \\
$k$-local Paulis & $3^{-k}$ (\autoref{prop:Pauli_commutation}) \\
Degree-$q$ Majoranas & $\Theta_q(n^{-q/2})$ (\autoref{thm:main_informal}) \\
All Paulis & $2^{-n}$ \cite[Lemma 5.8]{chen2022exponential} \\
\hline
\end{tabular}
\vspace{-0.3cm}
\caption{
The commutation index $\Delta(\mathcal{S})$ characterizes how non-commuting a set $\mathcal{S}$ of operators is. The commutation index reveals a key distinction between local spin operators and local fermionic operators: in the fermionic case, the commutation index decays polynomially with system size, while it is constant in the case of spins. The $\Theta_q$ notation assumes a fixed $q$ and growing $n$.
\label{tab:comm_index_results}}
\end{table}

The proof of \autoref{thm:main_informal} involves constructing the \emph{commutation graph} $G(\mathcal{S})$ whose vertices correspond to operators $\vA_i \in \mathcal{S}$ with edges between operators if and only if they anti-commute. The commutation index can be upper bounded by $\Delta \leq \vartheta(G(\mathcal{S})) / |G|$, where $\vartheta(G(\mathcal{S}))$ is the so-called \emph{Lovász theta function} of the commutation graph. The Lovász theta function can be efficiently computed via a semi-definite program~\cite{knuth1993sandwich}. For the SYK Hamiltonian, $G(\mathcal{S}^n_q)$ is the graph of a certain Johnson association scheme \cite{delsarte1973algebraic}.

In the course of writing our results we became aware of Ref.~\cite{linz2024systems}, which also establishes the necessary results on the Lovász theta function of Johnson association schemes. Our results use different proof techniques and determine the explicit $q$-dependence of the constant in Eq.~\eqref{eq:main_informal}, which was not derived in~\cite{linz2024systems}.

\section{Circuit lower bound for the SYK model} \label{sec:syk_nlts}

An almost direct consequence of a decaying commutation index is a lower bound on the complexity of any ansatz in constructing near-ground states.
For any random Hamiltonian $\vH=m^{-1/2}\sum_{i=1}^m g_i \vA_i$ with i.i.d.\ Gaussian coefficients $g_i$, the commutation index $\Delta(\{\vA_i\}_{i=1}^m)$ characterizes the maximum variance of the energy $\langle\psi|\vH|\psi\rangle$ for an arbitrary fixed state $|\psi\rangle$. Standard concentration bounds then imply that the probability a state $|\psi\rangle$ has energy $t$ is bounded as $\exp(-\Omega(t^2/\Delta))$.
This concentration is so strong that one can bound the maximum energy over extremely large sets of states (or $\epsilon$-nets of infinite sets) $\mathcal{S}$ via a simple union bound argument with high probability over the disorder. In particular, we obtain a lower bound $|\mathcal{S}| = \exp(\Omega(t^2/\Delta))$ on the cardinality of the class of ansatzes needed to achieve a given energy $t$.

Specializing to the SYK model via \autoref{thm:intro_NLTS}, we summarize the implications of this result for various classes of states $\mathcal{S}$ in \autoref{tab:SYK_complexity}. For instance, we show that all states that achieve a constant (i.e., $t=\Theta\left(1\right)$) approximation ratio with the SYK ground state energy have a quantum circuit depth of $\Omega_q(n^{q/2})$. In contrast, product states give constant factor approximations to the ground state energy for any local spin Hamiltonian (see \Cref{app:nlts} for a short proof). Our argument also extends to classical ansatzes. For instance, tensor network methods require a bond dimension that grows polynomially with $n$ to construct near-ground states \cite{schollwock2011density,banuls2023tensor}. Similarly, popular methods based on neural quantum states \cite{carleo2017solving,schmitt2020quantum,sharir2020deep,sharir2022neural,nomura2021dirac} need at least $\Omega(n^3)$ parameters to construct near-ground states for the standard $q=4$ SYK model, implying a bounded depth fully connected network must have layer width that grows as $\Omega(n^{3/2})$. 

\begin{table}
\centering
\begin{tabular}{lc}
\hline
\textbf{Ansatz} & \textbf{{Circuit complexity}$^*$} \\
\hline
Quantum circuit with $G$ gates & $G \geq \tilde{\Omega}_q(n^{q/2+1} t^2)$ \\
% \hline
MPS with bond dimension $\chi$ & $\chi \geq \Omega_q(n^{q/4 + 1/2} t)$ \\
% \hline
Neural network with $W$ parameters & $W \geq \Omega_q(n^{q/2 + 1} t^2)$ \\
\hline
% Gaussian State Energy Achievability & No Gaussian state achieves energy $t\sqrt{n}$ for constant $t$ with high probability \\
% \hline
\multicolumn{2}{l}{ \begin{footnotesize} $^*$min. complexity to achieve energy $t \lambda_{\max}(\vH^{\syk}_q)$ w.h.p. \end{footnotesize}}
\end{tabular}
\vspace{-0.3cm}
\caption{To achieve energy scaling as $t \lambda_{\max}(\vH^{\syk}_q)$ for the SYK Hamiltonian with high probability, ansatz complexity (e.g., circuit depth) must scale polynomially with $n$. See \Cref{app:nlts} for proofs. The $\Omega_q$ notations assume a fixed $q$ and growing $n$.}
\label{tab:SYK_complexity}
\end{table}

Our circuit lower bound is related to the study of  `no low-energy trivial states' (NLTS) Hamiltonians, whose existence was conjectured in \cite{freedman2013quantum} and resolved in \cite{anshu2023nlts,herasymenko2023fermionic}. However, the settings are not strictly comparable: our instances are random (average-case), whereas NLTS is formalized for worst-case bounded interaction instances of Hamiltonians. The randomness allows us to prove stronger statements in two ways.
%Although not strictly comparable, our results are stronger than the current NLTS theorems in two ways.
First, our circuit lower bounds hold for states at \emph{any} constant temperature, rather than for states below some energy threshold. Second, we can achieve arbitrary polynomial circuit depth lower bounds, whereas current constructions of NLTS only give a logarithmic depth lower bound. See \Cref{app:nlts} for more discussion.

\section{Annealed approximation for the SYK model} \label{sec:annealed}

The commutation index also has direct implications for the concentration of various physical properties of interest around their disordered expectation. One manifestation of this is in the relation between the \emph{quenched} and \emph{annealed} free energies:
\begin{equation}
    \underbrace{\frac{1}{n}\BE\ln{Z_\beta}}_{\text{quenched}} \quad \leq \quad \underbrace{\frac{1}{n}\ln{\BE Z_\beta}}_{\text{annealed}},
\end{equation}
where $Z_\beta$ is the partition function of the model $\sqrt{n}\bm{H}$ at an inverse temperature $\beta$. The quenched free energy assumes the disorder induced by the random couplings is fixed when averaging over thermal fluctuations; the annealed free energy treats these fluctuations on an equal footing. While the two quantities agree at high temperature, at low temperature the latter is incapable of accounting for frustration induced by the disorder of the random couplings which can induce a spin glass phase \cite{talagrand2000rigorous,parisi1979infinite}. Their disagreement is thus indicative of the presence of a spin glass phase (see \Cref{app:spin_glass_background}). Motivated by this we bound the difference in quenched and annealed free energies as a function of the temperature and the commutation index of the model:
\begin{equation}\label{eq:annealed_approx}
    \frac{1}{n}\BE\ln{Z_\beta} \leq \frac{1}{n}\ln{\BE Z_\beta}\leq\frac{1}{n}\BE\ln{Z_\beta}+4\beta^2\Delta.
\end{equation}
For the SYK model this directly implies \autoref{thm:annealed_low_temps}. Informally, this bound is due to controlling the growth of the moment generating function of $\ln\left(Z_\beta\right)-\mathbb{E}\left[\ln\left(Z_\beta\right)\right]$ using the commutation index $\Delta$. We formally prove Eq.~\eqref{eq:annealed_approx} in \Cref{app:concentration_and_annealed}. We there also prove concentration bounds for observable expectations as well as two-point correlators, again following from bounding how sensitive these quantities are when varying the disorder. We summarize some of these results when applied to the SYK model in \autoref{tab:concentrations_results}.

\begin{table}
\centering
\begin{tabular}{lc}
\hline
\textbf{Quantity $f$} & \textbf{Rate $K$} \\
\hline
$\lambda_{\text{max}}\left(\bm{H}_q^{\text{SYK}}\right)$ & $\Omega_q\left(n^{q/2}\right)$ \\
$\Tr\left(\bm{X}\bm{\rho}_\beta\right)$ & $\Omega_q\left(\beta^{-2}n^{q/2-1}\right)$ \\
$\Tr\left(\bm{H}_q^{\text{SYK}}\bm{\rho}_\beta\right)$ & $\Omega_q\left(\min\left(1,\beta^{-2}n^{-2}\right)n^{q/2}\right)^*$ \\
\hline
\multicolumn{2}{l}{ \begin{footnotesize} $^*$for $t$ order of $\left\lVert\bm{H}_q^{\text{SYK}}\right\rVert=\mathcal{O}(\sqrt{n})$ \end{footnotesize}}
\end{tabular}
\vspace{-0.3cm}
\caption{Concentration bounds for functions $f$ of the Hamiltonian around its mean, i.e., $\mathbb{P}\left[\left\lvert f-\mathbb{E}\left[f\right]\right\rvert\geq t\right]\leq 4\exp\left(-K t^2\right)$. $\lambda_{\max}$ denotes the largest eigenvalue and $\bm{X}$ is an arbitrary bounded operator. $\bm{\rho}_\beta$ is the thermal state of $\sqrt{n}\bm{H}$ at an inverse temperature $\beta$.}
\label{tab:concentrations_results}
\end{table}
\section{Lower Bound for the SYK Maximum Eigenvalue}

Finally, we show how commutation properties of local Hamiltonians can give rise to lower bounds on the maximum eigenvalue. In particular, for any random Hamiltonian $\bm{H}=m^{-1/2}\sum_{i=1}^m g_i\bm{A}_i$ where all $\vA_i^2 = \vI$, the scaling of the maximum eigenvalue is related to what we call the \emph{commutation degree} $h_{\text{comm}}\left(\left\{\bm{A}_i\right\}_{i=1}^m\right)$ of the operators comprising the Hamiltonian:
\begin{equation}
    h_{\text{comm}}\left(\left\{\bm{A}_i\right\}_{i=1}^m\right):=\frac{1}{2}\sup\limits_i\sum\limits_{j=1}^m\left\lVert\left[\bm{A}_i,\bm{A}_j\right]\right\rVert.
\end{equation}
The name `commutation degree' is derived from the fact that it is the maximal degree of the commutation graph associated with the set $\mathcal{S} = \{\bm{A}_1, \dots, \bm{A}_m\}$; see \Cref{app:commutation_graph}. The commutation degree can be interpreted as controlling the maximal amount of operator spreading of any $\bm{A}_i$ under the dynamics of $\bm{H}$. Using this intuition we are able to control how sensitive the expected partition function is to the inverse temperature:
\begin{equation}
    \frac{\partial }{\partial \beta} \BE \Tr [\e^{\beta \vH}]\geq\beta\left(1-\frac{c_1\beta^2 h_{\text{comm}}}{m}\right)\BE \Tr [\e^{\beta \vH}],
\end{equation}
where $c_1>0$ is an absolute constant. Using the lower bound of $\Tr [\e^{\beta \vH}] \leq \exp\left(\beta O(n) + \beta\lambda_{\text{max}}\left(\bm{H}\right)\right)$, we prove in \Cref{app:lowerbound} that, for all $\beta$,
\begin{equation}
    \mathbb{E}\exp\left(\beta\lambda_{\text{max}}\left(\bm{H}\right)\right)\geq\exp\L( \frac{\beta^2}{2} (1-\frac{c_1 \beta^2 h_{\text{comm}}}{2m} ) \R).
\end{equation}
Using the fact that the maximal eigenvalue concentrates (as in, e.g., \Cref{tab:concentrations_results}) and maximizing the bound over $\beta$ then implies:
\begin{equation}
    \BE \lambda_{\text{max}}(\vH) \ge \frac{\sqrt{m}}{4\sqrt{c_1 h_{\text{comm}}}}\left(1-16\Delta\right).
\end{equation}
By the same concentration results this bound also holds for $\lambda_{\text{max}}(\vH)$ with high probability over the disorder, not just in expectation. Intruigingly, the overall scaling depends only on the commutation degree $h_{\text{comm}}$. This quantity (when normalized by $m$) agrees to leading order for both the $q$-body quantum spin glass model $\bm{H}_q^{\text{SG}}$ and the SYK model $\bm{H}_q^{\text{SYK}}$. As long as $\Delta<1/16$---which is true for the former when $q\geq 3$, and always for the latter for sufficiently large $n$---this implies that for \emph{both} models:
\begin{equation}
    \lambda_{\text{max}}\left(\bm{H}_q^{\text{SYK}}\right)\geq\Omega\left(\frac{\sqrt{n}}{q}\right), \quad \lambda_{\text{max}}\left(\bm{H}_q^{\text{SG}}\right)\geq\Omega\left(\frac{\sqrt{n}}{q}\right).
\end{equation}
We here allow the locality $q$ to potentially grow with $n$ as $n\to\infty$. This strengthens the previously-known $\Omega\left(\sqrt{n}\right)$ scaling for both when $q$ is constant~\cite{hastings2022optimizing,herasymenko2023optimizing,anschuetz2024bounds}. For $q = \omega(\sqrt{n})$ it is known that both models exhibit a phase transition in their spectrums to a semicircle law~\cite{feng2019spectrum,erdHos2014phase} which is consistent with our results.

\begin{acknowledgments}
    E.R.A. is funded in part by the Walter Burke Institute for Theoretical Physics at Caltech. C.-F.C. is supported by a Simons-CIQC postdoctoral fellowship through NSF QLCI Grant No. 2016245. R.K. is funded by NSF grant CCF-2321079. We thank Chokri Manai, who pointed out an error in an earlier draft and shared his insights on its solution.
\end{acknowledgments}

\onecolumngrid
\clearpage
\bibliography{ref}

\appendix
\section{Background on related results for classical spin glass models} \label{app:spin_glass_background}

Spin glass models are a now-canonical object in the study of disordered systems in physics and mathematics. Properties of the energy landscape of this model feature phase transitions which govern the complexity of the near ground states and have various algorithmic and physical implications. We briefly summarize these important implications here, focusing on the (classical) Ising spin glass model. 

In the Ising spin model, configurations $\bm \sigma = (\sigma_i)_{i=1}^n$ are each a point in the space $\Sigma = \{-1,+1\}^n$ with energy given by the random Hamiltonian
\begin{equation} \label{eq:classical_ising_spin}
    H_C(\bm \sigma) = \frac{1}{\sqrt{\binom{n}{p}}}  \sum_{1 \leq i_1 < \cdots < i_p \leq n} J_{i_1 \dots i_p} \sigma_{i_1}  \cdots \sigma_{i_p},
\end{equation}
where $J_{i_1 \dots i_p} \sim \mathcal{N}(0,1)$ are coefficients drawn i.i.d.\ from the standard Normal distribution. This model can equivalently be viewed as a qubit or matrix model on the vector space $\mathbb{C}^{2^n}$. Denoting $\vZ_i$ as the Pauli Z matrix acting on qubit $i$, the Hamiltonian $\vH_C$ takes the form
\begin{equation}
    \vH_C = \frac{1}{\sqrt{\binom{n}{p}}}  \sum_{1 \leq i_1 < \cdots < i_p \leq n} J_{i_1 \dots i_p} \vZ_{i_1}  \cdots \vZ_{i_p}.
\end{equation}
Note that the eigenvectors of the above Hamiltonian map onto spins in Eq.~\ref{eq:classical_ising_spin} with energy given by the corresponding eigenvalue. For this model, the (quenched) free energy takes the form
\begin{equation}
    F(\beta) = n^{-1}\mathbb{E}\left[\log Z_\beta \right], \quad Z_\beta = \sum_{\bm \sigma \in \Sigma} \exp(\beta\sqrt{n} H_C\left(\bm \sigma\right)),
\end{equation}
with associated Gibbs measure
\begin{equation}
    \mu_\beta(A) = Z_\beta^{-1} \sum_{\bm \sigma \in A} \exp(-\beta \sqrt{n} H_C\left(\bm \sigma\right)).
\end{equation}

Given that for any $\bm \sigma$ the random variable $H_C(\bm \sigma)$ is distributed as Gaussian with unit variance, one can show that the limiting annealed free energy is
\begin{equation}
    \lim_{n \to \infty} n^{-1}\log \mathbb{E} \left[ Z_\beta \right] = \lim_{n \to \infty} n^{-1} \log\left(2^n \exp( \beta^2 n/2 \right)) = \log(2) + \beta^2/2.
\end{equation}
Explicitly calculating the quenched free energy is far more challenging. Nonetheless, Kirkpatrick and Thirumalai \cite{kirkpatrick1987p} predicted that below a critical temperature, the structure of the Gibbs distribution clusters into exponentially many disconnected clusters indicative of a so-called shattering phase of a spin model. Such a shattering phase is a characteristic of `glassiness' and a source of formal proofs for algorithmic hardness of finding near-ground states. We summarize these findings below.

Evidence of shattering was provided in \cite{talagrand2000rigorous} and only recently fully rigorously proven in \cite{gamarnik2023shattering} for the large $p$ limit. The phase transition coincides with the point where the quenched and annealed free energies fail to agree, formalized in \cite{talagrand2000rigorous} as:
\begin{equation} 
    \beta_p := \sup\left\{\beta:
    \limsup_{n\to\infty}\frac{\mathbb{E}[\log Z_\beta]}{n} = \log(2) + \frac{\beta^2}{2}\right\}.
\end{equation}
Talagrand shows in Theorem 1.1 of \cite{talagrand2000rigorous} that 
\begin{equation}
    (1-2^{-p})\sqrt{2 \log(2)} \leq \beta_p \leq \sqrt{2 \log(2)}.
\end{equation}
$\beta_p \to \sqrt{2 \log(2)}$ in the limit $p\to\infty$ and is notable for also being the transition point where the quenched and annealed free energies fail to agree for the Random Energy Model \cite{derrida1980random}. In the Random Energy Model, the $2^n$ different configurations each have an energy given by independent draws of a Gaussian random variable with variance $n$. In studying potential `glassiness' in SYK energy landscapes, \cite{gur2018does} study the distribution of low energy eigenvalues of SYK model. There, numerical evidence is presented that extremal eigenvalues of the SYK model have level repulsion, which is also observed in many random matrix theory eigenspectra but which is not a feature of the Random Energy Model.

Failure of the quenched free energy to agree with the annealed free energy is typically an indication that an exponentially small fraction of low energy states dominate contributions to the Gibbs distribution. Note that the quenched free energy for a particular draw of the coefficients is typically close to its expectation (i.e., self-averaging in the terminology of statistical mechanics), but this is not true of the partition function $Z_\beta$. Fluctuations in low energy states can cause $Z_\beta$ to oscillate significantly. When these outliers dominate the contribution to the Gibbs distribution, the annealed free energy can disagree with the quenched free energy.

The point $\beta_p$ intuitively indicates the transition into a regime where low-energy states are rare but nonetheless dominate the contribution to the Gibbs distribution. This fact has many implications. \cite{talagrand2000rigorous} shows that whether $\beta$ is greater or less than $\beta_p$ determines whether or not `overlaps' $R(\bm \sigma, \bm \sigma')$ of configurations $\bm \sigma, \bm \sigma'$ drawn from the Gibbs distribution converge to zero, where
\begin{equation}
    R(\bm \sigma, \bm \sigma') = n^{-1} \sum_i \sigma_i \sigma_i'.
\end{equation}
More explicitly, treating $R(\bm \sigma, \bm \sigma')$ as a random variable where $\bm \sigma, \bm \sigma'$ are drawn independently from the Gibbs measure at temperature $\beta$, Talagrand shows in Proposition 1.2 of \cite{talagrand2000rigorous} that
\begin{align}
    \beta < \beta_p & \implies \lim_{n \to \infty} \mathbb{E}\langle R(\bm \sigma, \bm \sigma')^2 \rangle_{\beta} = 0  \\
    \beta > \beta_p & \implies \exists \beta' < \beta, \quad \limsup_{N \to \infty} \mathbb{E}\langle R(\bm \sigma, \bm \sigma')^2 \rangle_{\beta'} > 0  \\
    \beta > \sqrt{2\log 2} & \implies \liminf_{N \to \infty} \mathbb{E}\langle R(\bm \sigma, \bm \sigma')^2 \rangle_{\beta} > 0,
\end{align}
where $\langle \cdot \rangle_\beta$ denotes the thermal or Gibbs average over independent replicas. $\liminf_{N \to \infty} \mathbb{E}\langle R(\bm \sigma, \bm \sigma')^2 \rangle_{\beta} > 0$ is the defining property of the glassy phase of a system, so this formally connects the disagreement of the quenched and annealed free energy with the onset of glassiness. Similar replica symmetry breaking behavior has also been proven to exist for Sherrington--Kirkpatrick models with a quantum transverse field (i.e., $\sum\limits_i\sigma_i^x$) of sufficiently low strength \cite{leschke2021existence,manai2024parisi}. Though we do not detail it here, for many glassy systems this transition is also associated with the onset of a clustering phenomenon in the landscape of low energy states known as the \emph{Overlap Gap Property} (OGP) \cite{gamarnik2021overlap}. This property is known to be the cause of the algorithmic hardness of optimizing glassy systems \cite{gamarnik2021circuit,huang2022tight,huang2023algorithmic}.

We now turn toward the SYK model. Let us denote the corresponding value of $\beta_p$ for the SYK model as
\begin{equation} 
    \beta_p^{\rm SYK} := \sup\left\{\beta:
    \limsup_{n\to\infty}\frac{\mathbb{E}[\log Z_\beta]}{n} = \lim_{n\to\infty} \frac{\log \mathbb{E}Z_\beta }{n}  \right\},
\end{equation}
where the free energy above is that of the SYK model and we assume the limit above exists for the annealed free energy.
We show in \autoref{thm:quench_annealed} that for $q \geq 4$, $\beta_p^{\rm SYK} $ must grow with $n$ and the transition governed by $\beta_p^{\rm SYK}$ cannot occur at any constant temperature. Insofar as the story from the spin glass setting agrees with that of the SYK model this would imply the lack of a transition into a clustered phase with topological barriers to optimization. Nonetheless, formalizing such notions of `glassiness' and the OGP for SYK models appear to be formidable tasks.

\section{Commutation index and the Lovász theta function}

\subsection{Commutation index}

In this section we study a property which quantifies the commutation structure of a set of operators. This allows us to discuss the commutation structure of local Majorana operators and how they differ from local Paulis.
\begin{defn} \label{def:comm_index}
For a set $\mathcal{S}$ of Hermitian operators, define their \emph{commutation index} by 
\begin{equation}
\Delta(\mathcal{S}) = \sup_{|\psi\rangle} \; \mathbb{E}_{\vA \in \mathcal{S}} \langle\psi|\vA|\psi\rangle^2.
\end{equation}
\end{defn}

The commutation index of the set of $k$-local Paulis is independent of system size $n$.

\begin{prop} \label{prop:Pauli_commutation}
Let $\mathcal{P}^n_k$ be the set of $k$-local $n$-qubit Paulis. When $2n+1\geq 3^k$, it holds:
\begin{equation}
\Delta(\mathcal{P}^n_k) = 3^{-k}.
\end{equation}
Moreover, the maximum is achieved by any product state.
\end{prop}
We also prove a slightly weaker bound that holds for any $k$. 
\begin{prop}\label{prop:anyk}
    Let $\mathcal{P}^n_k$ be the set of $k$-local $n$-qubit Paulis. Then, for any $k,n$, it holds:
\begin{equation}
\Delta(\mathcal{P}^n_k) \le \left(\frac{3}{2}\right)^{-k}.
\end{equation}
\end{prop}
See \Cref{app:pauli} for proofs of \autoref{prop:Pauli_commutation} and \autoref{prop:anyk}.

On the other hand, the commutation index of the set of degree-$q$ Majorana operators decays polynomially with system size.
\begin{defn}
The Majorana operators on $n$ fermionic modes are defined abstractly as $n$ operators $\{\gamma_1,\dots,\gamma_{n}\}$ which satisfy the relations
\begin{equation}
\gamma_a \gamma_b + \gamma_b \gamma_a = 2 \delta_{ab} \mathbbm{1}.
\end{equation}
A degree-$q$ Majorana operator is a degree-$q$ monomial in the Majorana operators.
\end{defn}

\begin{thm} \label{thm:Majorana_commutation}
Let $\mathcal{S}^n_q$ be the set of degree-$q$ Majorana operators on $n$ fermionic modes with $q$ even. Then
\begin{equation}
\left|\frac{\Delta(\mathcal{S}^n_q)}{{n/2 \choose q/2} \big/ {n \choose q}} - 1\right| \leq \mathcal{O}(n^{-1})
\end{equation}
for all $n$ sufficiently large, for each $q$.
\end{thm}
See \Cref{app:majorana} for a proof of \autoref{thm:Majorana_commutation}.

\subsection{Commutation graph} \label{app:commutation_graph}

Given a set $\mathcal{S}$ of Pauli or Majorana operators, we can study their commutation structure by encoding it into a graph $G$, which we call the \emph{commutation graph}.

\begin{defn} \label{def:commutation_graph}
\emph{(Commutation graph.)}
The \emph{commutation graph} $G(\mathcal{S})$ of a set $\mathcal{S}$ of Pauli or Majorana operators is defined as follows.
\begin{itemize}
    \item The vertices of $G(\mathcal{S})$ correspond to operators $\vA \in \mathcal{S}$.
    \item We include an edge between any two vertices whose operators \emph{anticommute}.
\end{itemize}
\end{defn}

We now introduce a key graph property which reveals the anticommutativity of the operators $\mathcal{S}$ through their commutation graph $G(\mathcal{S})$.

\begin{defn} \label{def:lovasz}
\emph{(Lovász theta function.)}
Let $G$ be a graph on $m$ vertices. The \emph{Lovász theta function} $\vartheta(G)$ is defined by the following semidefinite program of dimension $m$. Let $E$ denote the edges in the graph $G$, and $\mathbb{J}$ the all-ones matrix.
\begin{align}
    \max \ \{ \ &\Tr\left(\mathbb{J} \vX\right) \ , \ \vX \in \mathbb{R}^{m \times m} \nonumber\\
    &\text{s.t.} \ \vX \succeq 0 \ , \ \Tr\left(\vX\right) = 1 \ , \ \vX_{jl} = 0 \ \forall (j,l) \in E \ \}. \label{eq:theta_primal}
\end{align}
It has dual
\begin{align}
    \min \ \{ \ &\lambda \in \mathbb{R} \nonumber\\
    &\text{s.t.} \ \exists \ \vY \in \mathbb{R}^{m \times m} \ , \ \vY_{jj} = 1 \ \forall j \ , \ \vY_{jl} = 0 \ \forall (j,l) \notin E \ , \ \lambda \vY \succeq \mathbb{J} \ \}. \label{eq:theta_dual}
\end{align}
\end{defn}

For any graph $G$, the following chain of inequalities is known:
\begin{equation}
I(G) \leq \vartheta(G) \leq \chrom(\overline{G}),
\end{equation}
where $\overline{G}$ is the complement graph, $\chrom(\overline{G})$ is the chromatic number of $\overline{G}$, and $I(G)$ is the independence number of $G$. For example, see \cite{knuth1993sandwich}. Further, the commutation index $\Delta(\mathcal{S})$ is bounded by the Lovász theta function of $G(\mathcal{S})$. All together, these bounds are expressed in the following lemma.

\begin{lem} \label{lem:commutation_graph}
Let $\mathcal{S}$ be a set of Pauli operators
\begin{equation}
\Delta(\mathcal{S}) \leq \frac{1}{|\mathcal{S}|} \vartheta(G(\mathcal{S}))
\end{equation}
Further,
\begin{equation}
I(G(\mathcal{S})) \leq |\mathcal{S}| \cdot \Delta(\mathcal{S}) \leq \vartheta(G(\mathcal{S})) \leq \chrom(\overline{G}(\mathcal{S}))
\end{equation}
\end{lem}
\begin{proof}
The inequality $\Delta(\mathcal{S}) \leq \vartheta(G(\mathcal{S})) / |\mathcal{S}|$ has appeared previously in \cite[Equation 5]{de2023uncertainty}, \cite[Proposition 3]{xu2023bounding} and \cite{hastings2022optimizing}. The inequalities $I(G) \leq \vartheta(G) \leq \chrom(\overline{G})$ always hold for any graph $G$ \cite{knuth1993sandwich}. Finally, $\Delta(\mathcal{S}) \geq I(G(\mathcal{S})) / |\mathcal{S}|$ holds since we can choose $\rho$ in the definition of $\Delta(\mathcal{S})$ to be in the simultaneous eigenbasis of the independent set of operators.
\end{proof}

\subsection{Proof of \autoref{prop:Pauli_commutation}} \label{app:pauli}

First we aim to show $\mathbb{E}_{\vP \in \mathcal{P}^n_k} \langle\psi|\vP|\psi\rangle^2 = 3^{-k}
$ for any product state of single-qubit states $|\psi\rangle = |\psi_1\rangle \otimes \dots \otimes |\psi_n\rangle$. Denoting by $\mathcal{P}^S_k$ the set of Paulis on subsystem $S \subseteq [n]$, we have
\begin{equation}
\mathbb{E}_{\vP \in \mathcal{P}^n_k} \langle\psi|\vP|\psi\rangle^2 = \mathbb{E}_{S \subseteq [n] , |S| = k} \mathbb{E}_{\vP \in \mathcal{P}^S_k} \langle\psi|\vP|\psi\rangle^2
\end{equation}
By tracing out $[n] \setminus S$, it is sufficient to show
\begin{equation}
\mathbb{E}_{\vP \in \mathcal{P}^k_k} \langle\psi|\vP|\psi\rangle^2 = 3^{-k}
\end{equation}
for any product state $|\psi\rangle = |\psi_1\rangle \otimes \dots \otimes |\psi_k\rangle$. Since $|\mathcal{P}^k_k| = 3^k$, this is equivalent to
\begin{equation}
\sum_{\vP \in \mathcal{P}^k_k} \langle\psi|\vP|\psi\rangle^2 = 1
\end{equation}
But this holds since
\begin{align}
\sum_{\vP \in \mathcal{P}^k_k} \langle\psi|\vP|\psi\rangle^2 = \prod_{j=1}^k \Big(\sum_{\vP \in \{\sigma_X,\sigma_Y,\sigma_Z\}} \langle\psi_j|\vP|\psi_j\rangle^2\Big) = 1
\end{align}
using that the single-qubit states $|\psi_j\rangle$ are pure.

For the upper bound, we will invoke \autoref{lem:commutation_graph}. Recalling that $G(\mathcal{P}^n_k)$ is the commutation graph of $k$-local Paulis, it suffices to show that $\vartheta(G(\mathcal{P}^n_k)) \leq 3^{-k} \cdot |\mathcal{P}^n_k|$. For this purpose, we import a fact from \cite{knuth1993sandwich} using a proof technique similarly applied in \cite{anschuetz2024bounds}. A graph $G$ is \emph{vertex-symmetric} if for any two vertices $u,v$, there is an automorphism of $G$ taking $u$ to $v$.

\begin{fact}[\cite{knuth1993sandwich}]
If graph $G$ is vertex-symmetric, then
\begin{equation}
\vartheta(G) \cdot \vartheta(\overline{G}) = |G|
\end{equation}
where $\overline{G}$ is the complement graph and $|G|$ denotes the number of vertices in $G$.
\end{fact}

The commutation graph $G(\mathcal{P}^n_k)$ of $k$-local Paulis is vertex-symmetric. Thus to establish the upper bound $\vartheta(G(\mathcal{P}^n_k)) \leq 3^{-k} \cdot |\mathcal{P}^n_k|$, it suffices to show $\vartheta(\overline{G}(\mathcal{P}^n_k)) \geq 3^k$. Using the independence number inequality of \autoref{lem:commutation_graph}, it suffices to find an independent set in $\overline{G}(\mathcal{P}^n_k)$ of size at least $3^k$. This is equivalent to a \emph{clique} in $G(\mathcal{P}^n_k)$ of size at least $3^k$. In other words, we must exhibit a set of $3^k$ mutually anticommuting $k$-local Paulis. This can be done using the ternary tree embedding of \cite{vlasov2019clifford,jiang2020optimal}. Let $2n+1 = 3^k$. The ternary tree construction at depth $k$ embeds $2n+1$ mutually anticommuting operators into $n$ qubits where each anticommuting operator has locality $k$.

\subsection{Proof of \autoref{prop:anyk}}

For a fixed input state $|\psi\rangle$, we split the expectation by first sampling the choice of the subset $S\subset [n]$ and then the particular choice of Paulis.
    \begin{equation}
\mathbb{E}_{\vP \in \mathcal{P}^n_k} \langle\psi|\vP|\psi\rangle^2 = \mathbb{E}_{S \subseteq [n] , |S| = k} \mathbb{E}_{\vP \in \mathcal{P}^S_k} \langle\psi|\vP|\psi\rangle^2.
\end{equation}
Conditioned on each subset $S$, we compare the expectation over all $3^k$-many $k$-body Paulis with all $4^k$ possible Pauli strings acting on the subset (i.e., including identities):
\begin{align}
    \mathbb{E}_{\vP \in \mathcal{P}^S_k} \langle\psi|\vP|\psi\rangle^2 = \frac{1}{3^k}\sum_{\vP \in \mathcal{P}^S_k} \langle\psi|\vP|\psi\rangle^2 &\le \frac{1}{3^k}\sum_{\vP \in (\vI,\vsigma_x,\vsigma_y,\vsigma_z)^{\otimes k}} \langle\psi|\vP|\psi\rangle^2\tag{Including strings with identities}\\
    &\le \frac{1}{3^k} 4^k \bra{\psi}\tr_S[\ket{\psi}\bra{\psi}]\otimes \bm{\tau}_{S} \ket{\psi}\\
    &\le \frac{1}{3^k} 4^k \tr[ \tr_S[\ket{\psi}\bra{\psi}]^2\otimes\bm{\tau}_{S}^2]\\
    &\le \frac{1}{3^k} 4^k \frac{1}{2^k} = \left(\frac{2}{3}\right)^{k}.
\end{align}
The second line uses that the average over all Pauli strings acts by partial-tracing out $S$ and replacing with the maximally mixed state $\bm{\tau}_{S}$:
\begin{align}
    \frac{1}{4^k}\sum_{\vP \in \mathcal{P}^S_k} \vP[\vrho]\vP = \tr_S[\vrho] \otimes \bm{\tau}_{S}.
\end{align}
The last line uses that $\tr[\bm{\tau}_{S}^2] = 1/2^k$, and that the purity of the reduced state is bounded by $\tr[\vrho^2]\le 1.$ Take the expectation over subsets $\mathbb{E}_{S \subseteq [n], |S| = k}$ to conclude the proof.

\subsection{Proof of \autoref{thm:Majorana_commutation}} \label{app:majorana}

First we show the lower bound in \autoref{thm:Majorana_commutation}. We can find a set $S \subseteq \mathcal{S}^n_q$ of mutually commuting degree-$q$ Majoranas of size ${n/2 \choose q/2}$ by taking $(q/2)$-wise products of $\{i \gamma_1 \gamma_2, i \gamma_3 \gamma_4, \dots, 
i \gamma_{n-1} \gamma_{n}\}$. Let $\rho$ be the state which is maximally mixed within the simultaneous $+1$-eigenspace of the operators in $S$. Then $\sum_{\vA \in \mathcal{S}^n_q} \langle\psi| \vA |\psi\rangle = |S| = {n/2 \choose q/2}$ and $\Delta(\mathcal{S}^n_q) \geq {n/2 \choose q/2} \big/ {n \choose q}$. (Note the number of degree-$q$ monomials on $n$ Majoranas is $|\mathcal{S}^n_q| = {n \choose q}$.) The remainder of this section is devoted to showing the upper bound via the Lovász theta function. We aim to establish the following theorem.

\begin{thm} \label{conjecture}
Let $\mathcal{S}^n_q$ be the set of degree-$q$ Majorana operators on $n$ modes. Then
\begin{equation}
\vartheta(G(\mathcal{S}^n_q)) \leq {n/2 \choose q/2} + \mathcal{O}(e^{\mathcal{O}(q\log{q})} n^{q/2-1})
\end{equation}
for all $n$ sufficiently large, for each $q$.
\end{thm}

Noting that $|\mathcal{S}^n_q| = {n \choose q}$, the upper bound in \autoref{thm:Majorana_commutation} follows from combining \autoref{conjecture} and \autoref{lem:commutation_graph}. Thus it suffices to establish \autoref{conjecture}.

After completing our work, we became aware of Ref.~\cite{linz2024systems}, in which they establish $\vartheta(G(\mathcal{S}^n_q)) \leq {n/2 \choose q/2} + c(q) n^{q/2-1}$ for some function $c(q)$. \autoref{conjecture} is stronger, since it specifies the asymptotic dependence $c(q) = \mathcal{O}(e^{\mathcal{O}(q\log{q})})$. We give a self-contained proof of \autoref{conjecture}.

\bigskip

The \emph{Johnson association scheme} $\mathcal{J}_d(m,r)$ is the graph whose vertices correspond to subsets $S \subseteq [m]$ of size $|S| = r$, and $(S,T)$ forms an edge if $r - |S \cap T| = d$. Write $A^{m,r}_d$ for the adjacency matrix of the Johnson scheme $\mathcal{J}_d(m,r)$. The graph $G(\mathcal{S}^n_q)$ has adjacency matrix $A$ equal to
\begin{equation}
A = A^{n,q}_1 + A^{n,q}_3 + \dots + A^{n,q}_{q-1}
\end{equation}

We are interested in the Lovász theta function of $G(\mathcal{S}^n_q)$. The following three results advertised in \cite{hastings2022optimizing} reduce $\vartheta(G(\mathcal{S}^n_q))$ to a linear program involving Hahn polynomials.

\begin{lem} (\cite[p. 48]{delsarte1973algebraic})
The matrices $A^{m,r}_0,\dots,A^{m,r}_r$ are simultaneously diagonalizable, with eigenvalues given by the \emph{dual Hahn polynomials}:
\begin{align}
\text{spec}(A^{m,r}_d) &= \{\tilde{H}^{m,r}_d(x) : x = 0,\dots,r\} \\
\tilde{H}^{m,r}_d(x) &= \sum_{j=0}^d (-1)^{d-j} {r-j \choose d-j} {r-x \choose j} {m-r+j-x \choose j}
\end{align}
In particular, for all $d > 0$ the all-1’s vector is an eigenvector of $A^{m,r}_d$ of multiplicity 1 with eigenvalue $\tilde{H}^{m,r}_d(0) = {m-r \choose d} {r \choose d}$.
\end{lem}

\begin{lem}
In the dual formulation of the Lovász theta function in \autoref{def:lovasz} for the graph $G(\mathcal{S}^n_q)$ it suffices to minimize over matrices $\vY$ whose entries $\vY(S,T)$ depend only on $\text{dist}(S,T)$. Thus we can write
\begin{equation}
\vartheta(G(\mathcal{S}^n_q)) \ = \
\min \{ \lambda \
: \ \exists \ a_1,a_3,\dots,a_{q-1} \ \text{s.t.} \ \lambda \big(\mathbbm{1} + a_1 A^{n,q}_1 + a_3 A^{n,q}_3 + \dots + a_{q-1} A^{n,q}_{q-1}\big) \succeq \mathbb{J} \}.
\end{equation}
\end{lem}

\begin{cor} \label{cor:LP}
\begin{align}
\vartheta(G(\mathcal{S}^n_q)) \ = \
\min_{a_1,a_3,\dots,a_{q-1}} \{ &{n \choose q} \big/ \big(1 + p(0)\big) \
: \ p(1),\dots,p(r) \geq -1 \nonumber\\
&\text{where} \ p(x) = a_1 \tilde{H}^{n,q}_1(x) + a_3 \tilde{H}^{n,q}_3(x) + \dots + a_{q-1} \tilde{H}^{n,q}_{q-1}(x) \}.
\end{align}
\end{cor}

\bigskip

Our strategy to prove \autoref{conjecture} is as follows. We will examine the linear program of \autoref{cor:LP} in the large-$n$ limit. We will find that $p(0)$ has the correct limit for large $n$, given by the following lemma.
\begin{lem} \label{lem:p(0)}
$p(0) = \frac{2^{q/2} (q/2)!}{(q)!} n^{q/2} - c(q) n^{q/2-1}$ for some $c(q)$.
\end{lem}
This gives the correct large-$n$ limit for the Lovász theta function for constant $q$:
\begin{align}
{n \choose q} \big/ p(0) &= \left(\frac{1}{(q)!} \cdot n^{q} + \mathcal{O}(n^{q-1})\right) \cdot \left(\frac{2^{q/2} (q/2)!}{(q)!} n^{q/2} - \mathcal{O}(n^{q/2-1})\right)^{-1} = \frac{1}{(q/2)!} (n/2)^{q/2} + \mathcal{O}(n^{q/2-1})
\end{align}

Moreover, $p(0) / n^{q/2}$ will be a low-degree polynomial in $n^{-1}$.
\begin{lem} \label{lem:poly_degree}
$p(0) / n^{q/2}$ is a polynomial in $n^{-1}$ of degree $(q/2)(q/2-1)$.
\end{lem}

Finally, we will show a uniform bound on $|p(0) / n^{q/2}|$.
\begin{lem} \label{lem:uniform_bound}
$|p(0) / n^{q/2}| \leq e^{3{q/2}-3} q^{3{q/2}} / 4$ for all $n$.
\end{lem}

The polynomial method completes the argument by controlling the error for finite $n$. Markov's other inequality (\cite[Theroem 5.1.8]{polynomials_inequalities}) states the following (the particular bound on the $1/n$-values is worked out in, e.g.,~\cite{chen2024new,chen2024efficient}):
\begin{lem}[Markov's other inequality] \label{lem:markov_other_inequality}
For any polynomial $f(x)$ of degree $p$,
\begin{align}
\sup_{x\in [-1,1]} \labs{f'(x)} \le p^2\sup_{x\in [-1,1]} \labs{f(x)}.
\end{align}
Consequently, there is an absolute constant $c$ such that for any polynomial $f$ of degree $p$
\begin{align}
\labs{f(1/n)-f(0)} \le \frac{ cp^4}{n}\sup_{n' \ge 1}\labs{f(1/n')} \quad \text{for each integer}\quad n.
\end{align}
\end{lem}
Combining Lemmas \ref{lem:p(0)}, \ref{lem:poly_degree}, \ref{lem:uniform_bound} and \ref{lem:markov_other_inequality} completes the proof of \autoref{conjecture}.

\bigskip

For constant $r$, $x = 1,\dots,r$ and large $m$, the leading order term in the Hahn polynomial is
\begin{equation}
\tilde{H}^{m,r}_d(x) =
\begin{cases}
    \Theta(m^d) + \mathcal{O}(m^{d-1}) & d \leq r - x \\
    (-1)^{d+x-r} \Theta(m^{r-x}) + \mathcal{O}(m^{r-x-1}) & d > r - x
\end{cases}
\end{equation}
Since it will be important in our application later, let us be more specific about the coefficients in the cases $d = r - x - 1$ and $d = r - x + 1$. To leading order in $m$ with $r$ constant we have
\begin{equation}
\tilde{H}^{m,r}_{r-x-1}(x) = h^{(r)}(x) \cdot m^{r-x-1} + \mathcal{O}(m^{r-x-2}) \quad , \quad \tilde{H}^{m,r}_{r-x+1}(x) = -g^{(r)}(x) \cdot m^{r-x} + \mathcal{O}(m^{r-x-1})
\end{equation}
where
\begin{equation}
h^{(r)}(x) = \frac{r-x}{(r-x-1)!} \quad , \quad g^{(r)}(x) = \frac{x}{(r-x)!}
\end{equation}

Now let us examine the LP of \autoref{cor:LP} in the large-$n$ limit. Our strategy is to sequentially go through $x = q,\dots,1$ and ensure that $p(x) \geq -1$ for sufficiently large $n$ for each $x$. For odd $x$, the equation $p(x) \geq -1$ will automatically hold for sufficiently large $n$. Ensuring that $p(x) \geq -1$ eventually for even $x$ will require choosing $a_{q-x+1}$ as a function of $a_{q-x-1}$. All coefficients $a_1,a_3,\dots,a_{q-1}$ will be positive, and they will depend on $n$ with scaling $a_d = \Theta(n^{-(d-1)/2})$. Henceforth let us assume these properties, and we will see that they can be satisfied.

Let us first consider $p(q)$. For all $d$, $\tilde{H}^{n,q}_d(q)$ are negative and in fact independent of $n$. The $d=1$ term is equal to $-a_1 \cdot g^{(q)}(q)$, and the terms $d>1$ are subleading order in $n$ since they scale like $a_d = \Theta(n^{-(d-1)/2})$ by assumption. Thus the constraint $p(q) \geq -1$ is satisfied so long as
\begin{equation}
a_1 = \frac{C_1(n^{-1})}{g^{(q)}(q)}
\end{equation}
for some polynomial $C_1$ of degree $(q/2-1)$ with $C_1(0) = 1$ and $C_1(n^{-1}) \leq 1$ for all $n$.

For $p(q-1)$, all the Hahn polynomials appearing are positive to leading order in $n$, so $p(q-1)$ is large and positive eventually. The same holds for any odd value of $x$.

Now consider $p(q-2)$. The $d=1$ term is $a_1 \cdot h^{(q)}(q-2) \cdot n$ to leading order in $n$, and the $d=3$ term is $-a_3 \cdot g^{(q)}(q-2) \cdot n^2$ to leading order. The terms $d>3$ are subleading order in $n$ since they scale like $a_d \cdot \Theta(n^2) = \Theta(n^{-(d+1)/2})$. Thus we can satisfy the constraint $p(q-2) \geq -1$ by picking
\begin{align}
a_3 &= C_3(n^{-1}) \cdot \frac{h^{(q)}(q-2)}{g^{(q)}(q-2)} \cdot n^{-1} \cdot a_1 \\
&= C_3(n^{-1}) \cdot C_1(n^{-1}) \cdot \frac{h^{(q)}(q-2)}{g^{(q)}(q-2)} \cdot \frac{1}{g^{(q)}(q)} \cdot n^{-1}
\end{align}
for some polynomial $C_3$ of degree $(q/2-1)$ with $C_3(0) = 1$ and $C_3(n^{-1}) \leq 1$ for all $n$.

Continue like this for $p(q-3),\dots,p(1)$. For each even $x$, we will set
\begin{align}
a_{q-x+1} &= C_{q-x+1}(n^{-1}) \cdot \frac{h^{(q)}(x)}{g^{(q)}(x)} \cdot n^{-1} \cdot a_{q-x-1} \\
&= C_{q-x+1}(n^{-1}) \cdot \dots \cdot C_{1}(n^{-1}) \cdot \frac{h^{(q)}(x) \cdot h^{(q)}(x+2) \cdot \dots \cdot h^{(q)}(q-2)}{g^{(q)}(x) \cdot g^{(q)}(x+2) \cdot \dots \cdot g^{(q)}(q-2)} \cdot \frac{1}{g^{(q)}(q)} \cdot n^{-q/2 + (x/2)}
\end{align}
For each even $t$, $C_{q-t+1}$ is a polynomial of degree $(q/2-1)$ satisfying $C_{q-t+1}(0) = 1$ and $C_{q-t+1}(n^{-1}) \leq 1$ for all $n$. Notice that
\begin{equation}
\frac{h^{(q)}(t)}{g^{(q)}(t)} = \frac{(q-t)^2}{t}
\end{equation}
and $g^{(q)}(q) = q$, so defining
\begin{equation}
\hat{a}_{q-x+1} := \frac{(q-x)^2 (q-x-2)^2 \dots 2^2}{(q-2) (q-4) \dots x} \cdot \frac{1}{q}
\end{equation}
we can write
\begin{equation}
a_{q-x+1} = C_{q-x+1}(n^{-1}) \cdot \dots \cdot C_1(n^{-1}) \cdot \hat{a}_{q-x+1} \cdot n^{-q/2+(x/2)}
\end{equation}

Finally, let us look at $p(0)$. For large $n$, we have
\begin{align}
p(0) &= a_{q-1} \cdot h^{(q)}(0) \cdot n^{q-1} + \dots \\
&= C(n^{-1}) \cdot \frac{q}{(q-1)!} \cdot \hat{a}_{q-1} \cdot n^{q/2} - c(q) n^{q/2-1}
\end{align}
where $C = C_{q-1} \dots C_1$ is a polynomial of degree $(q/2)(q/2-1)$ satisfying $C(0) = 1$ and $C(n^{-1}) \leq 1$ for all $n$, and $c(q)$ is some function of $q$. (Note $h^{(q)}(0) = q / (q-1)!$.) The product in $\hat{a}_{q-1}$ telescopes to give
\begin{equation}
\hat{a}_{q-1} = \frac{1}{q} (q-2) (q-4) \dots 2
\end{equation}
so we get
\begin{equation}
p(0) = C(n^{-1}) \cdot \frac{1}{(q-1) (q-3) \dots 1} \cdot n^{q/2} - c(q) n^{q/2-1}
\end{equation}
as promised. This establishes \autoref{lem:p(0)}.

Let us now examine $p(0) / n^{q/2}$.
\begin{align}
p(0) / n^{q/2} &= \frac{1}{n^{q/2}} \Big(a_1 \tilde{H}^{n,q}_1(0) + a_3 \tilde{H}^{n,q}_3(0) + \dots + a_{q-1} \tilde{H}^{n,q}_{q-1}(0)\Big) \\
&= \hat{a}_1 {q \choose 1}{n-q \choose 1} \cdot n^{-q/2} \cdot C_1(n^{-1}) + \hat{a}_3 {q \choose 3}{n-q \choose 3} \cdot n^{-q/2-1} \cdot (C_1 C_3)(n^{-1}) + \\
&\qquad \dots + \hat{a}_{q-1} {q \choose q-1}{n-q \choose q-1} \cdot n^{-q+1} \cdot (C_1 \dots C_{q-1})(n^{-1})
\end{align}
recalling $\tilde{H}^{n,q}_d(0) = {q \choose d}{n-q \choose d}$. Note the coefficients $\hat{a}_d$ are independent of $n$. From this expression, we can readily see that $p(0) / n^{q/2}$ is a polynomial in $n$ of degree $(q/2)(q/2-1)$, establishing \autoref{lem:poly_degree}.

It remains to establish \autoref{lem:uniform_bound}. For all $d$, $(C_1\dots C_d)(n^{-1}) \leq 1$ and $\hat{a}_d \leq \hat{a}_{q-1} = 2^{q/2-1} (q/2-1)!$. Further, for all $d$
\begin{equation}
{q \choose d}{n-q \choose d} \leq \left(\frac{2e^q}{d^2}\right)^d n^d \leq (2e^q)^{q-1} n^d
\end{equation}
using the general bound ${m \choose r} \leq (em/r)^r$. Using these facts, we can bound
\begin{align}
\big|p(0)/n^{q/2}\big| &\leq 2^{q/2-1} (q/2-1)! \cdot (2e^q)^{q-1} \cdot \big(n^{-q/2+1} + n^{-q/2+2} + \dots + 1\big) \\
&\leq e^{3q/2-3} q^{3q/2} / 4
\end{align}
using $x! \leq x^{x+1}/e^{x-1}$ in the final step.

\subsection{Numerics on Lovász theta function of local Majorana operators}

In this section we present some numerics on the Lovász theta function of the commutation graph of degree-$q$ Majorana operators $\vartheta(G(\mathcal{S}^{n}_{q}))$.

\begin{table}[H]
\centering
\small  % Change font size to small
\begin{tabular}{|c|ccccc|ccccc|}
\hline
& \multicolumn{5}{c|}{$\vartheta(G(\mathcal{S}^{n}_{q}))$} & \multicolumn{5}{c|}{$n/2 \choose q/2$} \\
$n$ &	$q=2$ &	$q=4$ &	$q=6$ &	$q=8$ &	$q=10$ &	$q=2$ &	$q=4$ &	$q=6$ &	$q=8$ &	$q=10$  \\
\hline

2 &     1 &     &       &       &		     &       1 &		&		&		&		\\
4 &		2 &		1 &		&        &		     &       2 &		1 &		&		&		\\
6 &		3 &		3 &		1 &      &		     &       3 &		3 &		1 &		&		\\
8 &		4 &		14 &	4 &      1 &         &       4 &		6 &		4 &		1 &		\\
10 &		5 &		14.57 &	14.57 &   5 &     1 &         5 &		10 &	10 &	5 &		1 \\
12 &		6 &		15 &	52 &     15 &        6 &         6 &		15 &	20 &	15 &	6 \\
14 &		7 &		21 &	57.34 &  57.34 &     21 &        7 &		21 &	35 &	35 &	21 \\
16 &		8 &		28 &	64 &     198 &       64 &        8 &		28 &	56 &	70 &	56 \\
18 &		9 &		36 &	100.13 & 218.34 &    218.34 &    9 &		36 &	84 &	126 &	126 \\
20 &	10 &	45 &	153.11 & 251.22 &    787.17 &    10 &	45 &	120 &	210 &	252 \\
22 &    11 &    55 &    195.13 & 429.91 &   885.15 &    11 &    55 &    165 &   330 &   462 \\
24 &    12 &    66 &    236.42 & 759 &      982.84 &           12 &    66 &    220 &   495 &   792 \\
26 &    13 &    78 &    286 &    990.80 &   1757.0 &           13 &    78 &    286 &   715 &   1287 \\
28 &    14 &    91 &    364 &    1217.2 &   3260.2 &           14 &    91 &    364 &   1001 &  2002 \\
30 &    15 &    105 &   455 &    1444.2 &   4643.9 &           15 &    105 &   455 &   1365 &  3003 \\
32 &    16 &    120 &   560 &    1820.0 &   6040.7 &           16 &    120 &   560 &   1820 &  4368 \\
34 &    17 &    136 &   680 &    2423.3 &   7240.0 &           17 &    136 &   680 &   2380 &  6188 \\
36 &    18 &    153 &   816 &    3327.1 &   9269.4 &           18 &    153 &   816 &   3060 &  8568 \\
38 &    19 &    171 &   969 &    4512.8 &   12552 &           19 &    171 &   969 &   3876 &  11628 \\
40 &    20 &    190 &   1140 &   6022.1 &   17230 &           20 &    190 &   1140 &  4845 &  15504 \\
\hline
\end{tabular}
\caption{Numerical comparison of the Lovász theta function $\vartheta(G(\mathcal{S}^{n}_{q}))$ versus $n/2 \choose q/2$. They are exactly equal for very small values of $n$, and also appear to be exactly equal for sufficiently large values of $n$ for each $q$. For example at $q=4$, which corresponds to the standard SYK-4 model, it appears that $\vartheta(G(\mathcal{S}^{n}_{4})) = {n/2 \choose 2}$ for all even values of $n$ apart from $n = 8$ and $n = 10$.}
\label{tab:my_label}
\end{table}

\begin{figure}[H]
\centering
  \includegraphics[]{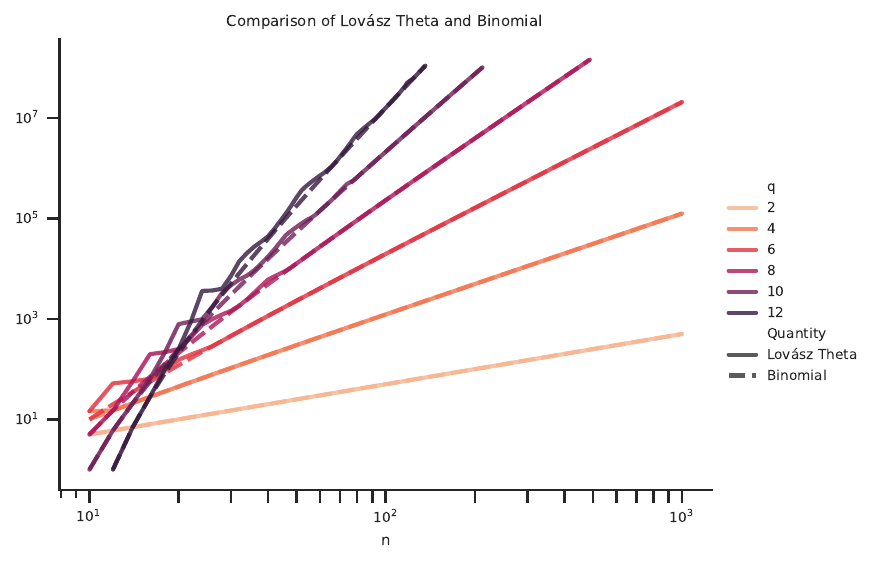}
  \caption{Log-log plot of $\vartheta(G(\mathcal{S}^{n}_{q}))$ versus $n/2 \choose q/2$. $\vartheta(G(\mathcal{S}^{n}_{q}))$ fluctuates for small $n$, but for sufficiently large $n$ it behaves the same as $n/2 \choose q/2$.}
  \label{fig:Lovasz}
\
\end{figure}

\subsection{Alternative definition of commutation index}

\begin{lem}
    When the left and right test vectors are different, we still have
    \begin{equation}
        \Delta(\mathcal{S}) \le \sup_{\norm{\ket{u}}=\norm{\ket{v}}=1} \mathbb{E}_{\vA \in \mathcal{S}} \labs{\bra{u}\vA\ket{v}}^2 = \sup_{\vO}  \frac{\mathbb{E}_{\vA \in \mathcal{S}}\labs{\tr[\vA \vO]}^2}{\norm{\vO}_1^2} \le 16 \Delta(\mathcal{S}).
    \end{equation}
\end{lem}
\begin{proof} 
The first inequality is trivial. The middle equality follows from the first by taking the singular value decomposition of $\vO$. To prove the final inequality, for every $u,v$, define polarizations 
\begin{align}
    \ket{s}:=\frac{\ket{u}+s\ket{v}}{2} \quad \text{where}\quad s = \pm 1, \pm i.
\end{align}

Then, for each $\vA$, 
\begin{align}
\labs{\bra{u}\vA \ket{v}}^2 &= \labs{4\BE_s s\bra{s}\vA\ket{s}}^2\\
&\le 16 \BE_{s}\labs{\bra{s}\vA\ket{s}}^2.
\end{align}
Hence,
\begin{align}
    \sup_{\norm{\ket{u}}=\norm{\ket{v}}=1} \mathbb{E}_{\vA \in \mathcal{S}}\labs{\bra{u}\vA \ket{v}}^2 &\le 16 \sup_{\norm{\ket{u}}=\norm{\ket{v}}=1} \mathbb{E}_{\vA \in \mathcal{S}} \BE_{s}\labs{\bra{s}\vA\ket{s}}^2\\
    &= 16 \sup_{\norm{\ket{u}}=\norm{\ket{v}}=1} \BE_{s}\mathbb{E}_{\vA \in \mathcal{S}} \labs{\bra{s}\vA\ket{s}}^2\\
    &\le 16 \Delta(\mathcal{S}).
\end{align}
\end{proof}

\section{Annealed approximation and concentration results}
\label{app:concentration_and_annealed}

Consider the model
\begin{align}
\vH = \frac{1}{\sqrt{m}} \sum_{i=1}^{m} g_i \vA_i,
\end{align}
where $g_i \sim_{i.i.d.} \mathcal{N}(0,1)$ are standard independent Gaussians and $\vA_i$ are deterministic matrices. The \emph{Gibbs state} $\vrho_\beta$ at inverse temperature $\beta$ is defined by
\begin{equation}
\vrho_\beta = \frac{\e^{-\beta \sqrt{n} \vH}}{Z_\beta} \quad , \quad Z_\beta = \Tr\left(\e^{-\beta \sqrt{n} \vH}\right),
\end{equation}
where $Z_\beta$ is called the \emph{partition function} at inverse temperature $\beta$. The factor $\sqrt{n}$ ensures that the free energy is extensive and scales proportionally to $n$. 

In this section we show that the commutation index of the terms $\vA_i$ has an important effect on the concentration properties of the random model $\vH$. Denote the commutation index by
\begin{equation}
\Delta := \Delta(\{\vA_i\}_{i=1}^m).
\end{equation}
Recall that this quantity characterizes the variance of the energy with respect to a fixed state $\vrho$:
\begin{align}
\sup_{\vrho} \BE_{\vH}\labs{\Tr(\vH \vrho)}^2 = \sup_{\vrho} \frac{1}{m} \sum_{i=1}^m (\Tr(\vA_i \vrho))^2 = \Delta.
\end{align}
The value of $\Delta$ has implications for relations between the normalized quenched and annealed free energies:
\begin{equation}
    \underbrace{\frac{1}{n}\BE\ln{Z_\beta}}_{\text{quenched}} \quad vs. \quad \underbrace{\frac{1}{n}\ln{\BE Z_\beta}}_{\text{annealed}}.
\end{equation}
The quenched free energy is physical but hard to calculate while the annealed free energy is much easier to calculate but nonphysical. The first result is that this variance quantity controls the difference between the two.
\begin{thm}[Quenched and annealed free energy]\label{thm:quench_annealed}
\begin{align}
n^{-1}\BE \ln{Z_\beta} \le n^{-1}\ln \BE Z_\beta \le n^{-1}\BE \ln{Z_\beta} + 4\beta^2 \Delta.
\end{align}
\end{thm}
The first inequality always holds, and the non-trivial part is the second inequality.
Therefore, a small variance $\Delta \ll \beta^{-2} $ means that the annealed free energy well-approximates the quenched free energy, which indicates the absence of spin glass order~\cite{Swingle20_SK_SYK}.
The next three results give concentration of expectation values, energy, and two-point correlators of the thermal state of $\vH$. Two-point correlators are of special interest in the study of the SYK model \cite{Swingle20_SK_SYK,gur2018does,kitaev2018soft,Maldacena_rmkSYK}. Concentration results for Lipschitz bounded functions of the spectrum of the SYK model have also been established in \cite{feng2020spectrumIII}. 

\begin{thm}[Concentration of expectation values]\label{thm:concentration_expectation}
For any fixed bounded Hermitian operator $\vX$,
\begin{align}
\mathbb{P}\big(\big|\Tr\left(\vX \vrho_\beta\right) - \mathbb{E}{\Tr\left(\vX \vrho_\beta\right)}\big| \geq t\big) \leq 2 e^{-t^2 / (18 \beta^2 ||\vX||^2 \Delta)}.
\end{align}
\end{thm}

\begin{thm}[Concentration of energy]\label{thm:concentration_energy}
\begin{align}
\mathbb{P}\big(\big|\Tr\left(\vH \vrho_\beta\right) - \mathbb{E}{\Tr\left(\vH \vrho_\beta\right)}\big| \geq t\big) \leq 4 \exp\left(-\frac{1}{2\Delta}\left(\sqrt{\frac{t^2}{12\beta^2 n} + \alpha^2} - \alpha\right)\right)
\end{align}
where $\alpha = \frac{1}{2}\big(1/(4\beta^2 n)+\mathbb{E}[\lambda_{\max}(\vH)]^2\big)$. 
\end{thm}

\begin{thm}[Concentration of two-point correlators]\label{thm:concentration_two_point}
For any fixed bounded operators $\vX$ and $\vY$, denoting
\begin{equation}
    \vY\left(\tau\right)=\exp\left(i \sqrt{n}\vH\tau\right)\vY\exp\left(-i \sqrt{n}\vH\tau\right)
\end{equation}
for any $\tau\in\mathbb{R}$, we have:
\begin{align}
\mathbb{P}\big(\frac{1}{2}\big|\Tr\left(\vX\vY\left(\tau\right)\vrho_\beta\right) - \mathbb{E}{\Tr\left(\vX\vY\left(\tau\right)\vrho_\beta\right)}\pm\text{h.c.}\big| \geq t\big) \leq 2 e^{-t^2 / (6n(5\beta^2+16\tau^2)\left\lVert\bm{X}\right\rVert^2\left\lVert\bm{Y}\right\rVert^2\Delta)}.
\end{align}
\end{thm}

Instantiating for example Theorems \ref{thm:quench_annealed}, \ref{thm:concentration_expectation}, \ref{thm:concentration_energy}, \ref{thm:concentration_two_point} with the upper bound on the commutation index of Majoranas in \autoref{thm:Majorana_commutation} yields the following results for the SYK model.

\begin{cor} \label{thm:syk_annealed}
\emph{(SYK model is annealed.)}
For the SYK model $\vH^{\syk}_q$ where $q$ is even,
\begin{align}
&\frac{1}{n}\BE \ln{Z_\beta} \le \frac{1}{n}\ln \BE Z_\beta \le \frac{1}{n}\BE \ln{Z_\beta} + \mathcal{O}_q(\beta^2 n^{-q/2}), \\
&\mathbb{P}\big(\big|\Tr\left(\vX \vrho_\beta\right) - \mathbb{E}{\Tr\left(\vX \vrho_\beta\right)}\big| \geq t\big) \leq 2 e^{-\Omega_q(\beta^{-2} n^{q/2-1} t^2)}, \\
&\mathbb{P}\big(\big|\Tr\left(\vH^{\syk}_q \vrho_\beta\right) - \mathbb{E}{\Tr\left(\vH^{\syk}_q \vrho_\beta\right)}\big| \geq t\big) \leq \begin{cases}
    4e^{-\Omega_q\left(\beta^{-1}n^{q/2-1/2} t\right)} & t=\Omega\left(1+\beta n\right) \\
    4e^{-\Omega_q\left(\min\left(1,\beta^{-2}n^{-2}\right)n^{q/2}t^2\right)} & \text{otherwise}
\end{cases}, \\
&\mathbb{P}\big(\frac{1}{2}\big|\Tr\left(\vX\vY\left(\tau\right)\vrho_\beta\right) - \mathbb{E}{\Tr\left(\vX\vY\left(\tau\right)\vrho_\beta\right)}\pm\text{h.c.}\big| \geq t\big) \leq 2 e^{-\Omega_q\left(\min\left(\beta^{-2},\tau^{-2}\right)n^{q/2-1}t^2\right)}.
\end{align}
$\vX$ and $\vY$ are any fixed bounded operators.
\end{cor}

Importantly, the above result shows that for the standard SYK model with $q=4$, the quenched free energy in the limit of $n\to\infty$ always equal its annealed approximation for physical temperatures where $\beta=\Theta(\sqrt{n})$. This stands in stark contrast with spin glasses where a transition occurs for some critical temperature $\beta_p$ into a clustered or `glassy' phase (see \Cref{app:spin_glass_background}).

The remainder of this appendix is concerned with establishing Theorems \ref{thm:quench_annealed}, \ref{thm:concentration_expectation}, \ref{thm:concentration_energy}, \ref{thm:concentration_two_point} for concentration of various observables and free energies.

\subsection{Preliminaries}

Let us first state a useful fact about Lipschitz functions of Gaussian variables.
\begin{fact}[Gaussian concentration of Lipschitz functions, Theorem~2.26 of \cite{Wainwright2019}] \label{fact:gaussian_lipschitz}
    Let $\bm{g} = (g_1,\dots,g_m)$ be i.i.d. standard Gaussian variables, and $f:\mathbb{R}^m \rightarrow \mathbb{R}$ $L$-Lipschitz. Then for any $t \geq 0$:
    \begin{equation}
        \mathbb{P}\big(|f(\bm{g}) - \mathbb{E}{f(\bm{g})}| \geq t\big) \leq 2 e^{-t^2 / (2L^2)}
    \end{equation}
\end{fact}

We also state a useful fact on the concentration of the operator norm of random matrices of the form of $\vH$.
\begin{fact}[Concentration of the maximal eigenvalue {~\cite[Corollary 4.14]{Bandeira2021MatrixCI}}]\label{fact:concentration_max_eigen}
    Let $\lambda_{\text{max}}(\vH)$ be the maximal eigenvalue of $\vH$. We have:
    \begin{align}
    \Pr\L(\lambda_{\text{max}}(\vH) - \BE \lambda_{\text{max}}(\vH) \ge t \R)    \le \exp\L(-\frac{t^2}{2\Delta}\R).
    \end{align}
\end{fact}
    
In the course of proving our results we will also use an equivalent formulation of \autoref{fact:gaussian_lipschitz} that follows from its sub-Gaussianity~\cite{rigollet2015course,vershynin2018high,wainwright2019high}.
\begin{lem}[Sub-Gaussian MGF bound, Lemma~1.5 of Ref.~\cite{rigollet2015course}] \label{lem:eig_mgf_bound}
    Given a random variable $X$ with sub-Gaussian concentration bound
    \begin{equation}
        \mathbb{P}\big(|X - \mathbb{E}{X}| \geq t\big) \leq 2 e^{-t^2 / (2\sigma^2)},
    \end{equation}
    it holds that
    \begin{equation}
        \mathbb{E}{\left[\exp\left(t\left(X-\mathbb{E}{X}\right)\right)\right]}\leq\exp\left(4\sigma^2 t^2\right).
    \end{equation}
\end{lem}

\subsection{Proof of \autoref{thm:quench_annealed}}

% \begin{proof}
    We directly compute the derivatives of  $\ln{Z_\beta}$ with respect to each Gaussian $g_i$ 
    \begin{align}
        \partial_{g_i} \ln{Z_\beta} &= \frac{1}{Z_\beta}  \Tr[\partial_{g_i} e^{\beta \sqrt{n} \vH}] \\
        &= \frac{1}{Z_\beta} \Tr\left[ \beta \sqrt{\frac{n}{m}}  \int_{0}^1 e^{\beta \sqrt{n} \vH (1-s)} A_i e^{\beta \sqrt{n} \vH} ds \right] \tag*{(Derivative of matrix exponential~\cite{10.1063/1.1705306})} \\
        &= \beta \sqrt{\frac{n}{m}}  \Tr[A_i \vrho_\beta]. \tag*{(Cyclic property of trace)}
    \end{align}
    Therefore, the Lipschitz constant $L$ of $\ln{Z_\beta}$ with respect to the disorder has the gradient bound: 
    \begin{equation}
        L^2 \leq  \frac{\beta^2 n}{m} \sum_{i=1}^{m} \Tr[A_i \vrho_\beta]^2 \leq \beta^2 n \Delta.
    \end{equation}
    Now we can bound
    \begin{equation}
        \frac{\BE\left[ Z_\beta\right]}{ \exp(\BE[\ln{Z_\beta}])} = \BE\left[ \exp( \ln{Z_\beta} - \BE[\ln{Z_\beta}]) \right]  \leq \exp(4\beta^2 n \Delta).
    \end{equation}
    The inequality uses \autoref{fact:gaussian_lipschitz} and \autoref{lem:eig_mgf_bound} with $t=1$. Taking logarithms and rearrange to obtain
    \begin{equation}
        \frac{1}{n}\log \BE[Z_\beta] \leq \frac{1}{n} \BE[ \ln{Z_\beta} ] + 4\beta^2 \Delta,
    \end{equation}
    as stated.
% \end{proof}

\subsection{Proof of \autoref{thm:concentration_expectation}}

The result once again follows from a Lipschitz bound. We use the well-known expression for the derivative of a matrix exponential~\cite{10.1063/1.1705306}:
\begin{equation}
    \partial_{g_i}\exp\left(\bm{H}\right)=\int\limits_0^1\exp\left(t\bm{H}\right)\left(\partial_{g_i}\bm{H}\right)\exp\left(\left(1-t\right)\bm{H}\right)\,dt.
\end{equation}
From the chain rule we then have:
\begin{equation}
    \partial_{g_i}\Tr\left(\bm{\rho}_\beta\vX\right)=\frac{\beta \sqrt{n} }{\sqrt{m}}\left(Z_\beta^{-1}\Tr\left(\vX\int\limits_0^1\exp\left(t\beta \sqrt{n} \bm{H}\right)\bm{A}_j\exp\left(\left(1-t\right)\beta \sqrt{n} \bm{H}\right)\,dt\right) - \Tr{\left(\vX\bm{\rho}_\beta\right)} \Tr{\left(\bm{A}_j\bm{\rho}_\beta\right)}\right).
\end{equation}

Consider now the operator:
\begin{equation}
    \bm{\sigma}_\beta\equiv Z_\beta^{-1}\int\limits_0^1\exp\left(t\beta \sqrt{n} \bm{H}\right)\vX\exp\left(\left(1-t\right)\beta \sqrt{n} \bm{H}\right)\,dt.
\end{equation}
We can check that $\bm{\sigma}_\beta$ has trace norm bounded by $\left\lVert\vX\right\rVert$. Denoting by $\Sigma_i\left(\cdot\right)$ the $i$th singular value of $\cdot$ in nonincreasing order, we have by the majorization inequality \cite{bhatia1997matrix}:
\begin{equation}
    \sum\limits_i\Sigma_i\left(\bm{A}\bm{B}\right)\leq\sum\limits_i\Sigma_i\left(\bm{A}\right)\Sigma_i\left(\bm{B}\right)
\end{equation}
that
\begin{equation}
\begin{split}
    \sum\limits_i \Sigma_i\left( \exp\left(t\beta \sqrt{n} \bm{H}\right)\vX\exp\left(\left(1-t\right)\beta \sqrt{n} \bm{H}\right)\right) 
    &\leq \left\lVert\bm{X}\right\rVert\sum\limits_i\Sigma_i\left(\exp\left(t\beta \sqrt{n} \bm{H}\right)\right)\Sigma_i\left(\exp\left(\left(1-t\right)\beta \sqrt{n} \bm{H}\right)\right) \\ 
    &\leq\left\lVert\bm{X}\right\rVert Z_\beta.
\end{split}
\end{equation}
This implies that $\bm{\sigma}_\beta$ has trace norm bounded by $\left\lVert\bm{X}\right\rVert$. However, $\bm{\sigma}_\beta$ is Hermitian but not necessarily positive semidefinite. We instead consider $\bm{\sigma}_\beta$ as the difference of two positive semidefinite matrices:%\AC{You have assumed that $\sigma$ is hermitian?}\era{yeah, I guess I implicitly assumed $\bm{X}$ is Hermitian (after that, $\bm{\sigma}$ is Hermitian by $t\leftrightarrow 1-t$ symmetry of the integral). I can make this explicit in the theorem statement}
\begin{equation}
    \bm{\sigma}_\beta=\bm{\sigma}_\beta^+-\bm{\sigma}_\beta^-,
\end{equation}
where each of $\bm{\sigma}_\beta^\pm$ is a (unnormalized) quantum state. By the cyclic property of the trace we can then write:
\begin{equation} \label{eq:exp_deriv}
    \partial_{g_i}\Tr\left(\bm{\rho}_\beta\vX\right) = \frac{\beta \sqrt{n} }{\sqrt{m}}\left(\Tr\left(\bm{\sigma}_\beta^+\bm{A}_j\right)-\Tr\left(\bm{\sigma}_\beta^-\bm{A}_j\right)-\Tr\left(\vX\bm{\rho}_\beta\right)\Tr\left(\bm{A}_j\bm{\rho}_\beta\right)\right).
\end{equation}
We thus have:
\begin{align}
    \big|\big|\bm{\nabla}_{\bm{g}}\Tr\left(\bm{\rho}_\beta\vX\right)\big|\big|_2^2 &= \frac{\beta^2 n}{m}\sum\limits_j\left(\Tr\left(\bm{\sigma}_\beta^+\bm{A}_j\right)-\Tr\left(\bm{\sigma}_\beta^-\bm{A}_j\right)-\Tr\left(\vX\bm{\rho}_\beta\right)\Tr\left(\bm{A}_j\bm{\rho}_\beta\right)\right)^2\\
    &\leq \frac{3\beta^2n}{m} \sum\limits_j\left(\big(\Tr{\bm{\sigma}_\beta^+\bm{A}_j}\big)^2 + \big(\Tr{\bm{\sigma}_\beta^-\bm{A}_j}\big)^2 + \big(\Tr{\vX\bm{\rho}_\beta}\big)^2 \big(\Tr{\bm{A}_j\bm{\rho}_\beta}\big)^2\right)\\
    &\leq 9 \beta^2 n ||\vX||^2 \Delta.
\end{align}
The result then follows from \autoref{fact:gaussian_lipschitz}.

\subsection{Proof of \autoref{thm:concentration_energy}}

We would like an analog of Eq.~\eqref{eq:exp_deriv} where the observable is $\vH$. Notice this is $g_i$-dependent and commutes with $\bm{\rho}_\beta$. We get:
\begin{equation}
    \partial_{g_i}\Tr\left(\bm{\rho}_\beta\bm{H}\right)=\frac{1}{\sqrt{m}}\Tr\left(\bm{\rho}_\beta\bm{A}_j\right)+\frac{\beta \sqrt{n} }{\sqrt{m}}\left(\Tr\left(\bm{\rho}_\beta\bm{H}\bm{A}_j\right)-\Tr\left(\bm{H}\bm{\rho}_\beta\right)\Tr\left(\bm{A}_j\bm{\rho}_\beta\right)\right).
\end{equation}
Let $\lambda_{\text{max}}(\vH)$ denote the maximal eigenenergy of $\vH$, and let $\mathcal{G}_s$ be the set of coefficients $\bm{g}$ where $\lambda_{\text{max}}\left(\bm{H}\right)\leq s+\mathbb{E}{\lambda_{\text{max}}\left(\bm{H}\right)}$. For $\bm{g}\in\mathcal{G}_s$:
\begin{align}\label{eq:energy_lip_bound}
    \big|\big|\bm{\nabla}_{\bm{g}}\Tr\left(\bm{\rho}_\beta\bm{H}\right)\big|\big|_2^2 &= \frac{1}{m}\sum\limits_j\left(\Tr\left(\bm{\rho}_\beta\bm{A}_j\right) + \beta \sqrt{n}  \Tr\left(\bm{\rho}_\beta\bm{H}\bm{A}_j\right) - \beta \sqrt{n}  \Tr\left(\bm{H}\bm{\rho}_\beta\right)\Tr\left(\bm{A}_j\bm{\rho}_\beta\right)\right)^2 \\
    &\leq \frac{3}{m} \sum_j \left(\big(\Tr{\bm{\rho}_\beta \vA_i}\big)^2 + \beta^2 n \big(\Tr{\bm{\rho}_\beta \vH \vA_i}\big)^2 + \beta^2 n \big(\Tr{\bm{\rho}_\beta \vH}\big)^2 \big(\Tr{\bm{\rho}_\beta \vA_i}\big)^2\right) \\
    &\leq 3\Delta \big(1 + 2 \beta^2 n ||\vH||^2\big) \\
    &\leq 3\Delta \big(1 + 2 \beta^2 n \left(s+\mathbb{E}{\lambda_{\text{max}}\left(\bm{H}\right)}\right)^2\big).
\end{align}
By \autoref{fact:concentration_max_eigen}, $\mathbb{P}\left[\bm{g}\not\in\mathcal{G}_s\right]\leq 2\exp\left(-\frac{s^2}{2\Delta}\right)$. Furthermore, $\mathcal{G}_s$ is a convex set, and thus there exists a function $\hat{\Tr}\left(\bm{\rho}_\beta\bm{H}\right)$ that agrees with $\Tr{\bm{\rho}_\beta\bm{H}}$ on $\mathcal{G}_s$ yet has the same Lipschitz bound of Eq.~\eqref{eq:energy_lip_bound} on the full domain. We thus calculate:
\begin{align}
    \mathbb{P}\left[\left\lvert\Tr\left(\bm{H}\bm{\rho}_\beta\right) - \mathbb{E}\left[\Tr\left(\bm{H} \bm{\rho}_\beta\right)\right]\right\rvert \geq t\right]&\leq\inf_s\mathbb{P}\left[\left\lvert\Tr\left(\bm{H}\bm{\rho}_\beta\right) - \mathbb{E}\left[\Tr\left(\bm{H} \bm{\rho}_\beta\right)\right]\right\rvert \geq t\wedge\bm{g}\in\mathcal{G}_s\right]+2\exp\left(-\frac{s^2}{2\Delta}\right)\\
    &=\inf_s\mathbb{P}\left[\left\lvert\hat{\Tr}\left(\bm{H}\bm{\rho}_\beta\right) - \mathbb{E}\left[\hat{\Tr}\left(\bm{H} \bm{\rho}_\beta\right)\right]\right\rvert \geq t\wedge\bm{g}\in\mathcal{G}_s\right]+2\exp\left(-\frac{s^2}{2\Delta}\right)\\
    &\leq\inf_s\mathbb{P}\left[\left\lvert\hat{\Tr}\left(\bm{H}\bm{\rho}_\beta\right) - \mathbb{E}\left[\hat{\Tr}\left(\bm{H} \bm{\rho}_\beta\right)\right]\right\rvert \geq t\right]+2\exp\left(-\frac{s^2}{2\Delta}\right)\\
    &\leq\inf_s 2 \exp\left(-\frac{t^2}{6\Delta \big(1 + 2 \beta^2 n \left(s+\mathbb{E}\left[\lambda_{\text{max}}\left(\bm{H}\right)\right]\right)^2\big)}\right) + 2\exp\left(-\frac{s^2}{2\Delta}\right)\\
    &\leq\inf_s 2 \exp\left(-\frac{t^2}{6\Delta \big(1 + 4 \beta^2 n \left(s^2+\mathbb{E}\left[\lambda_{\text{max}}\left(\bm{H}\right)\right]^2\right)\big)}\right) + 2\exp\left(-\frac{s^2}{2\Delta}\right)\\
    &\leq\inf_s 2 \exp\left(-\frac{t^2}{24 \beta^2 \Delta n \big(s^2 + 1/(4\beta^2 n)+\mathbb{E}[\lambda_{\text{max}}(\vH)]^2\big)}\right) + 2\exp\left(-\frac{s^2}{2\Delta}\right).
\end{align}
Setting $s^2 = \sqrt{t^2/(12\beta^2 n) + \alpha^2} - \alpha$ where $\alpha = \frac{1}{2}\left(1/(4\beta^2 n)+\mathbb{E}[\lambda_{\text{max}}(\vH)]^2\right)$ gives the desired result.

\subsection{Proof of \autoref{thm:concentration_two_point}}

Completely analogously to the proof of \autoref{thm:concentration_expectation} we have:
\begin{equation}\label{eq:two_pt_deriv}
    \partial_{g_i}\Tr\left(\bm{\rho}_\beta\vX\vY\left(\tau\right)\right)=\frac{\beta\sqrt{n}}{\sqrt{m}}\left(\Tr\left(\bm{\sigma}_\beta\bm{A}_j\right)-\Tr\left(\vX\vY\left(\tau\right)\bm{\rho}_\beta\right)\Tr\left(\bm{A}_j\bm{\rho}_\beta\right)\right)+\Tr\left(\bm{\rho}_\beta\vX\partial_{g_i}\vY\left(\tau\right)\right),
\end{equation}
where:
\begin{equation}
    \bm{\sigma}_\beta=Z_\beta^{-1}\int\limits_0^1\exp\left(t\beta\sqrt{n}\bm{H}\right)\vX\vY\left(\tau\right)\exp\left(\left(1-t\right)\beta\sqrt{n}\bm{H}\right)\,dt.
\end{equation}
We now focus on the final term of Eq.~\eqref{eq:two_pt_deriv}. We have:
\begin{equation}
    \partial_{g_i}\vY\left(\tau\right)=\frac{i\tau\sqrt{n}}{\sqrt{m}}\left(\int\limits_0^1\exp\left(i\tau\sqrt{n} t\bm{H}\right)\bm{A}_j\exp\left(i \tau\sqrt{n}\left(1-t\right)\bm{H}\right)\,dt\right)\vY\exp\left(-i\tau\sqrt{n}\vH\right)+\text{h.c.}=\frac{i\tau\sqrt{n}}{\sqrt{m}}\left[\bm{\tilde{A}}_{j|\tau},\vY\left(\tau\right)\right],
\end{equation}
where $\bm{\tilde{A}}_{j|\tau}$ is the Hermitian, time-averaged operator:
\begin{equation}
    \bm{\tilde{A}}_{j|\tau}=\int\limits_0^1\exp\left(i\tau\sqrt{n} t\bm{H}\right)\bm{A}_j\exp\left(-i \tau\sqrt{n} t\bm{H}\right)\,dt=\frac{1}{\tau}\int\limits_0^\tau\exp\left(it\sqrt{n}\bm{H}\right)\bm{A}_j\exp\left(-i t\sqrt{n}\bm{H}\right)\,dt.
\end{equation}
We have: 
\begin{align}
    \frac{1}{4}\big|\big|&\bm{\nabla}_{\bm{g}}\Tr\left(\bm{\rho}_\beta\vX\vY\left(\tau\right)\right)\pm\text{h.c.}\big|\big|_2^2\\
    &=\frac{n}{4m}\sum\limits_j\left\lvert\beta\Tr\left(\bm{\sigma}_\beta\bm{A}_j\right)-\beta\Tr\left(\vX\vY\left(\tau\right)\bm{\rho}_\beta\right)\Tr\left(\bm{A}_j\bm{\rho}_\beta\right)+i\tau\Tr{\bm{\rho}_\beta\vX\left[\bm{\tilde{A}}_{j|\tau},\vY\left(\tau\right)\right]}\pm\text{h.c.}\right\rvert^2\\
    &\leq\frac{3n}{4m}\sum\limits_j\left(\left\lvert\beta\Tr\left(\bm{\sigma}_\beta\bm{A}_j\right)\pm\text{h.c.}\right\rvert^2+\left\lvert\beta\Tr\left(\vX\vY\left(\tau\right)\bm{\rho}_\beta\right)\Tr\left(\bm{A}_j\bm{\rho}_\beta\right)\pm\text{h.c.}\right\rvert^2+\left\lvert\tau\Tr\left(\bm{\rho}_\beta\vX\left[\bm{\tilde{A}}_{j|\tau},\vY\left(\tau\right)\right]\right)\pm\text{h.c.}\right\rvert^2\right).
\end{align}
To proceed, consider (for instance):
\begin{equation}
    \Tr\left(\left(\bm{Y}\left(\tau\right)\bm{\rho}_\beta\bm{X}+\text{h.c.}\right)\bm{\tilde{A}}_{j|\tau}\right)\equiv\Tr\left(\bm{\mu}_{\beta,\tau}\bm{A}_j\right).
\end{equation}
$\bm{\mu}_{\beta,\tau}$ is Hermitian by definition and has trace norm bounded by $2\left\lVert\bm{X}\right\rVert\left\lVert\bm{Y}\right\rVert$ by the matrix H\"{o}lder inequality (or alternatively, an analog of the singular value majorization argument used in the proof of \autoref{thm:concentration_expectation}). Once again we can consider $\bm{\mu}_{\beta,\tau}$ as the difference of two positive semidefinite matrices:
\begin{equation}
    \bm{\mu}_{\beta,\tau}=\bm{\mu}_{\beta,\tau}^+-\bm{\mu}_{\beta,\tau}^-.
\end{equation}
Doing this for all terms yields:
\begin{align}
    \frac{1}{4}\big|\big|&\bm{\nabla}_{\bm{g}}\Tr\left(\bm{\rho}_\beta\vX\vY\left(\tau\right)\right)\pm\text{h.c.}\big|\big|_2^2\\
    &\leq\frac{3n}{4m}\sum\limits_j\left(\left\lvert\beta\Tr\left(\bm{\sigma}_\beta\bm{A}_j\right)\pm\text{h.c.}\right\rvert^2+\left\lvert\beta\Tr\left(\vX\vY\left(\tau\right)\bm{\rho}_\beta\right)\Tr\left(\bm{A}_j\bm{\rho}_\beta\right)\pm\text{h.c.}\right\rvert^2+\left\lvert\tau\Tr\left(\bm{\rho}_\beta\vX\left[\bm{\tilde{A}}_{j|\tau},\vY\left(\tau\right)\right]\right)\pm\text{h.c.}\right\rvert^2\right)\\
    &\leq\frac{3n}{4}\left(16\beta^2\left\lVert\bm{X}\right\rVert^2\left\lVert\bm{Y}\right\rVert^2\Delta+4\beta^2\left\lVert\bm{X}\right\rVert^2\left\lVert\bm{Y}\right\rVert^2\Delta+64\tau^2\left\lVert\bm{X}\right\rVert^2\left\lVert\bm{Y}\right\rVert^2\Delta\right)\\
    &=3n\left(5\beta^2+16\tau^2\right)\left\lVert\bm{X}\right\rVert^2\left\lVert\bm{Y}\right\rVert^2\Delta.
\end{align}

\section{Lower bound on SYK optimum} \label{app:lowerbound}

Consider a Hamiltonian weighted by i.i.d.\ Gaussian coefficients 
\begin{align}
    \vH = \frac{1}{\sqrt{m}}\sum_{i=1}^m g_i \vA_i \quad \text{where}\quad \BE g^2_i =1
\end{align}
where we here assume that all $\vA_i^2 = \norm{\vA_i}^2\vI$. We show a lower bound for the maximal eigenvalue for such a Hamiltonian. 
\begin{thm}[Lower bound on the maximal eigenvalue]\label{thm:bounding_maximal} There is an absolute constant $c_1$ such that
\begin{align}
    \BE \lambda_{\text{max}}(\vH) \ge \frac{\sqrt{m}}{4\sqrt{c_1 h_{\text{comm}}}}\left(h_{\text{glo}}^2-16\Delta\right).
\end{align}
\end{thm}
The commutation index $\Delta = \Delta(\{\vA_j\}_{j=1}^m)$ is defined in \autoref{def:comm_index}. We have also defined the \emph{commutation degree}
\begin{equation}
    h_{\text{comm}} := \frac{1}{2}\sup_{i} \sum_{j=1}^m \frac{\norm{\vA_j}\cdot\norm{[\vA_i,\vA_j]}}{\norm{\vA_i}}
\end{equation}
and the \emph{global norm}
\begin{equation}
    h_{\text{glo}}:=\sqrt{\frac{1}{m}\sum\limits_{i=1}^m\norm{\vA_i}^2}.
\end{equation}
Our results in the main text are only reported in the case all $\norm{\vA_i}=1$ for simplicity. Note that here, both $h_{comm}$ and $h_{glo}$ feature a quadratic sum  (instead of a linear sum) due to the randomness of the Hamiltonian. 

This immediately gives lower bounds for the SYK maximal eigenvalue.
\begin{cor} \label{cor:lambda_max_syk}
With high probability over the disorder, the maximum eigenvalue of the SYK model is
\begin{equation} \label{eq:lambda_max_syk}
\lambda_{\max}(\vH^{\syk}_q) = \Omega(\sqrt{n} / q).
\end{equation}
\end{cor}
\begin{proof}
We have $h_{\text{comm}}=\mathcal{O}(q\binom{n-1}{q-1})$ and $m=\binom{n}{q}$ for SYK. $\Delta=o\left(1\right)$ by \autoref{conjecture}. The result then follows from \autoref{thm:bounding_maximal} and the concentration of the maximal eigenvalue (\autoref{fact:concentration_max_eigen}).
\end{proof}

Our theorem also lower bounds the maximal eigenvalue of a $k$-local quantum spin glass:
\begin{equation}\label{eq:qsg}
    \bm{H}_k^{\text{SG}}=\frac{1}{\sqrt{3^k\binom{n}{k}}}\sum\limits_{k\text{-local }\bm{\sigma}}g_{\bm{\sigma}}\bm{\sigma}.
\end{equation}
\begin{cor} \label{cor:lambda_max_sg}
With high probability over the disorder, the maximum eigenvalue of the $k$-local quantum spin glass model is
\begin{equation} \label{eq:lambda_max_sg}
\lambda_{\max}(\vH_k^{\text{SG}}) = \Omega(\sqrt{n} / k)
\end{equation}
when $k\geq 3$.
\end{cor}
\begin{proof}
We have $h_{\text{comm}}=\mathcal{O}(k3^{k-1}\binom{n-1}{k-1})$ and $m=3^k\binom{n}{k}$ for the quantum spin glass described in Eq.~\eqref{eq:qsg}. $\Delta$ is also less than $1/16$ when $k\geq 3$ by \autoref{prop:Pauli_commutation}. The result then follows from \autoref{thm:bounding_maximal} and the concentration of the maximal eigenvalue (\autoref{fact:concentration_max_eigen}).
\end{proof}

\bigskip

The strategy to prove \autoref{thm:bounding_maximal} is to lower bound the optimum by calculating the exponential 
\begin{align}
     \e^{\beta \BE\lambda_{\text{max}}(\vH)} &\approx \BE \e^{\beta \lambda_{\text{max}}(\vH)} \tag*{(Concentration of the maximal eigenvalue:~\autoref{fact:concentration_max_eigen})}\\
     &\ge \BE \frac{1}{N} \sum_i\e^{\beta \lambda_{i}(\vH)} = \BE \btr [\e^{\beta \vH}],
\end{align}
where we use $\btr\left[\cdot\right]:=\frac{1}{N}\Tr\left[\cdot\right]$ to denote the normalized trace. We begin by lower bounding the right-hand side.
\begin{lem}[Lower bounds on the exponential]\label{lem:lower_bound_exponential}
There is an absolute constant $c_1$ such that for each $\beta,$
\begin{align}\label{eq:exp_bound}
    \BE\btr[\e^{\beta \vH}] \ge  \exp\L( \frac{\beta^2 h_{\text{glo}}^2}{2} (1-\frac{c_1\beta^2 h_{\text{comm}}}{2m} ) \R).
\end{align}
\end{lem}

In proving Lemma~\ref{lem:lower_bound_exponential} we will use the below two facts.
\begin{fact}[Integration by parts]\label{lem:byparts}
For standard Gaussian random variable $g$ and a function $f:\BR\rightarrow \BR$ whose derivative is absolutely integrable w.r.t. the Gaussian measure, we have that
\begin{align}
    \BE[g f(g) ] =  \BE[f'(g)].
\end{align}
\end{fact}
\begin{fact}[Multivariate H{\"o}lder for random matrices{e.g., ~\cite[Fact A.1]{chen2023sparse}}]\label{fact:products_holder}

For any family $(\vX_1,\dots,  \vX_{k})$ of square random matrices, possibly statistically dependent,
the product satisfies the trace inequality
\begin{align}
  \Expect  \btr \labs{\prod_{i=1}^{k} \vX_i} 
  = \vertiii{\prod_{i=1}^{k} \vX_i }_1 
  \le \prod_{i=1}^{k} \vertiii{\vX_i}_{p_i} \quad \text{whenever} \quad \sum_{i=1}^{k} \frac{1}{p_i} = 1
  \quad\text{and}\quad p_i \geq 0,
\end{align}
where
\begin{align}
    \vertiii{\vX}_p:=\L(\Expect  \btr [\labs{\vX}^p] \R)^{1/p}.
\end{align}
\end{fact}

\begin{proof}[Proof of Lemma~\ref{lem:lower_bound_exponential}]
Take derivative w.r.t.\ $\beta$:
\begin{align}
    \frac{\partial }{\partial \beta} \BE \btr [\e^{\beta \vH}]&=\BE \btr [\vH\e^{\beta \vH}]\\
    &= \frac{1}{\sqrt{m}}\sum_{i=1}^m \BE\btr [g_i \vA_i\e^{\beta \vH}]\\
    &= \frac{1}{\sqrt{m}}\sum_{i=1}^m \BE\btr [\vA_i \partial_i \e^{\beta \vH}] \tag*{(Integration by parts:~\autoref{lem:byparts})}\\
    &= \frac{\beta}{m} \sum_{i=1}^m \int_0^1 \BE\btr [\vA_i \e^{\beta \vH s} \vA_i \e^{\beta \vH (1-s)}] ds \tag*{(Derivative of matrix exponential~\cite{10.1063/1.1705306})}\\
    &= \frac{\beta}{m} \sum_{i=1}^m \int_0^1 \BE\btr [\vA^2_i \e^{\beta \vH} ]ds +\frac{\beta^2}{m} \sum_{i=1}^m \int_{s_1+s_2 +s_3 =1} \BE\btr [\vA_i \e^{\beta \vH s_3}  [\vH,\vA_i] \e^{\beta \vH s_2} \e^{\beta \vH s_1}] ds.
\end{align}
The last line ``swaps'' the $\vA_i$ through the exponential, resulting in errors written as commutator, as seen from 
\begin{align}
    \e^{\vX} \vY -\vY \e^{\vX} = \int_0^1 \e^{\vX s}[\vX,\vY]\e^{\vX (1-s)} ds
\end{align}
and setting $\vX = s \vH.$ The intuition is that the commutator would be small due to the locality of $\vA_i$. We further expand the second term
\begin{align}
    &\int_{s_1+s_2 +s_3 =1} \BE\btr [\vA_i \e^{\beta \vH s_3}  [\vH,\vA_i] \e^{\beta \vH (s_1+s_2)}  ] ds \\
    &=\frac{1}{\sqrt{m}}\sum_{j=1}^m \int_{s_1+s_2 +s_3 =1} \BE\btr [\vA_i \e^{\beta \vH s_3}  [g_j\vA_j,\vA_i] \e^{\beta \vH (s_1+s_2)}  ] ds\\
    &= \frac{\beta}{m} \sum_{j=1}^m \int_{s_1+s_2 +s_3+ s_4 =1} \BE\btr [\vA_i \e^{\beta \vH s_4} \vA_j \e^{\beta \vH s_3}  [\vA_j,\vA_i] \e^{\beta \vH (s_1+s_2)}] ds +(\text{other insertions of~} \vA_j ) \tag*{(Integration by parts:~\autoref{lem:byparts})}.
\end{align}
Thus,
\begin{align}
    &\labs{ \frac{\beta^2}{m} \sum_{i=1}^m \int_{s_1+s_2 +s_3 =1} \BE\btr [\vA_i \e^{\beta \vH s_3}  [\vH,\vA_i] \e^{\beta \vH s_2} \e^{\beta \vH s_1}] ds} \\
    &\le (\text{const.}) \frac{\beta^3}{m^2} \sum_{i,j=1}^m \norm{\vA_i} \norm{\vA_j }\cdot \norm{[\vA_i,\vA_j]} \cdot \BE\btr [\e^{\beta \vH}]\tag*{(\autoref{fact:products_holder} and $\e^{\vH} \succ 0$)}\\
    &= (\text{const.}) \frac{\beta^3}{m^2} \sum_{i=1}^m \norm{\vA_i}^2 \sum_{j=1}^m \norm{\vA_j }\frac{\norm{[\vA_i,\vA_j]}}{\norm{\vA_i}} \cdot \BE\btr [\e^{\beta \vH}]\\
    &\le c_1 \frac{\beta^3}{m} h_{\text{glo}}^2 h_{\text{comm}} \cdot \BE\btr [\e^{\beta \vH}]\tag*{(By definitions of $h_{\text{glo}}$ and $h_{\text{comm}}$)}.
\end{align}
The second inequality uses a multivariate H\"{o}lder's inequality for random matrices (\autoref{fact:products_holder}), for moment parameters $p = 1/s$ for each $\e^{\beta\vH s}$, and $p = \infty$ for each $\vA_i, \vA_j$ and $[\vA_i,\vA_j].$ Indeed, $\vertiii{ \e^{\beta \vH s }}_{1/s}= (\BE \btr[\labs{\e^{\beta \vH}}])^{s}=(\BE \btr[\e^{\beta \vH}])^{s}.$ Also, the ``$(\text{const.})$'' notation absorbs the absolute numerical constants arising from the insertions of $\vA_j$ and the integration over $\sum_{\ell} s_{\ell} =1.$ Thus, defining $f\left(\beta\right):=\BE \btr [\e^{\beta \vH}]$,
\begin{align}
    \frac{\partial }{\partial \beta } f(\beta) &\ge \beta h_{\text{glo}}^2 f(\beta) - c_1 \frac{\beta^3}{m} h_{\text{glo}}^2 h_{\text{comm}} f(\beta)\\
    \implies\quad f(\beta) &\ge \exp\L( \int_0^{\beta} \beta' h_{\text{glo}}^2 (1- c_1 \frac{\beta^{'2}}{m} h_{\text{comm}}) d\beta' \R) f(0) \tag*{(Gronwall's differential inequality)}\\
    &\ge \exp\L( \frac{\beta^2 h_{\text{glo}}^2}{2}  (1-\frac{c_1\beta^2 h_{\text{comm}}}{2m} ) \R)\tag*{(By the initial condition $f(0)= 1$)},
\end{align}
which concludes the proof.
\end{proof}
One immediate corollary of this lemma is the following, which follows from evaluating Eq.~\eqref{eq:exp_bound} at $\beta=\sqrt{\frac{m}{c_1 h_{\text{comm}}}}$.
\begin{cor}[A good $\beta$]\label{cor:good_beta}
\begin{align}
     \BE\btr[ \e^{\beta_{\text{max}}\vH}] \ge \exp\L( \frac{h_{\text{glo}}^2 m}{4c_1 h_{\text{comm}}}\R)\quad \text{at}\quad \beta_{\text{max}}:= \sqrt{\frac{m}{c_1 h_{\text{comm}}}}.
\end{align}    
\end{cor}

With the preliminaries in place we are now able to prove \autoref{thm:bounding_maximal}. We will also use eigenvalue concentration bounds proved in \Cref{app:concentration_and_annealed}.
\begin{proof}[Proof of~\autoref{thm:bounding_maximal}]
We have the lower bound:
\begin{align}   \ln \BE \e^{\beta_{\text{max}} \lambda_{\text{max}}(\vH)} &\ge \ln \BE\btr[ \e^{\beta_{\text{max}}\vH}]\\
&\ge \beta_{\text{max}}\cdot \frac{h_{\text{glo}}^2\sqrt{m}}{4\sqrt{c_1 h_{\text{comm}}}}
\tag*{(\autoref{cor:good_beta})}.
\end{align}
Also, we move the expectation to the exponent by concentration of the maximal eigenvalue
\begin{align}
     \BE \e^{ \beta \lambda_{\text{max}}(\vH)} &= \e^{\beta \BE \lambda_{\text{max}}(\vH)} \BE \e^{\beta \lambda_{\text{max}}(\vH) - \beta \BE \lambda_{\text{max}}(\vH)}\\
     &\le \e^{\beta \BE \lambda_{\text{max}}(\vH)} \e^{4\beta^2\Delta}\tag*{(\autoref{fact:concentration_max_eigen} and \autoref{lem:eig_mgf_bound})}.
\end{align}
Rearrange and take the logarithm to conclude the proof.
\end{proof}
\section{Circuit lower bound for SYK model} \label{app:nlts}

In this section we show that low-energy states of random strongly interacting fermionic Hamiltonians have high circuit complexity. In particular, we show a circuit lower bound on the low energy states of the the SYK model. It was previously described in Eq.~\eqref{eq:syk_model}, but we repeat its definition here for convenience.

\begin{defn} \label{def:SYK}
Let $\mathcal{S}^n_q$ denote the set of degree-$q$ Majorana operators on $n$ fermionic modes. The $\syk_q$ model is a random ensemble of Hamiltonians defined by
\begin{equation}
\vH^{\syk}_q = \frac{1}{\sqrt{n \choose q}} \sum_{\vA \in \mathcal{S}^n_q} g_{\vA} \vA \quad , \quad g_{\vA} \sim_{i.i.d.} \mathcal{N}(0,1).
\end{equation}
\end{defn}

\begin{thm} \label{thm:fermion_NLTS}
\emph{(SYK model low-energy states have high circuit complexity.)}
Let $\operatorname{circ}(G)$ denote the set of unitaries generated by quantum circuits with at most $G$ gates each taken from a finite universal set of $2$-local unitary gates. Fix an arbitrary initial state $\ket{\phi}$.
With high probability, for any even $q \geq 2$, it holds that the minimum circuit complexity to construct a state achieving at least $t \sqrt{n}$ on $\vH^{\syk}_q$ is at least
\begin{equation}
    \min \left\{G: \;\exists U \in \operatorname{circ}(G), \; \bra{\phi}U^\dagger\vH^{\syk}_q U\ket{\phi} \geq t \sqrt{n} \right\} = \tilde{\Omega}_q(n^{(q/2)+1} t^2).
\end{equation}
\end{thm}

Meanwhile, we can recall \autoref{cor:lambda_max_syk} from \Cref{app:lowerbound}, which gives a lower bound $\lambda_{\max}(\vH^{\syk}_q) = \Omega_q(\sqrt{n})$ on the maximum eigenvalue of SYK.

The proof of \autoref{thm:fermion_NLTS} will proceed via a concentration argument, followed by a union bound over the circuit family. This resembles the circuit lower bound of \cite[Appendix D]{dalzell2023sparse}. We establish the necessary concentration now, which relies crucially on \autoref{thm:Majorana_commutation} concerning the commutation index of low-degree Majorana operators.

\begin{lem} \label{lem:syk_concentration}
Fix any state $|\psi\rangle$. The energy $\langle\psi|\vH^{\syk}_q|\psi\rangle$ sharply concentrates:
\begin{equation}
\mathbb{P}\big(\langle\psi|\vH^{\syk}_q|\psi\rangle \geq t\big) \leq \exp(- \Omega_q(n^{q/2} t^2)).
\end{equation}
\end{lem}
\begin{proof}
Since a sum of Gaussians is Gaussian, we have
\begin{equation}
\mathbb{P}\big(\langle\psi|\vH^{\syk}_q|\psi\rangle \geq t\big) \leq \exp(-t^2 / 2 \sigma^2)
\end{equation}
where
\begin{align}
\sigma^2 &= \frac{1}{{n \choose q}} \sum_{\vA \in \mathcal{S}^n_q} \langle\psi|\vA|\psi\rangle^2 \leq \Delta(\mathcal{S}^n_q) \leq \frac{(q)!}{2^{q/2} (q/2)!} n^{-q/2} + \mathcal{O}(e^{\mathcal{O}(q\log{q})} n^{-(q/2)-1}) = \mathcal{O}_q(n^{-q/2}).
\end{align}
Here $\Delta$ is from \autoref{def:comm_index}, and we used the upper bound of \autoref{thm:Majorana_commutation}.
\end{proof}

\begin{proof}[Proof of \autoref{thm:fermion_NLTS}.]
Let $M$ be the number of gates in the universal gate set. Then the number of circuits in $\operatorname{circ}(G)$ is at most 
\begin{equation}
    |\operatorname{circ}(G)| \leq \left( M {n \choose 2} \right)^G = \exp( \mathcal{O}(G \log(n))).
\end{equation}
Performing a union bound on \autoref{lem:syk_concentration} yields
\begin{equation}
\begin{split}
    \mathbb{P}\left[ \max_{U \in \operatorname{circ}(G)} \bra{\phi}U^\dagger\vH^{\syk}_q U\ket{\phi} \geq t \sqrt{n} \right] \leq  \exp\left(- \Omega_q(n^{q/2+1} t^2) + \mathcal{O}(G \log(n)) \right).
\end{split}
\end{equation}
For this to be non-vanishing in $n$, it requires
\begin{equation}
G = \Omega_q(n^{q/2+1} t^2 / \log{n} ).
\end{equation}
\end{proof}

\begin{rmk}
    The proof above can be extended to gates with continuous parameters by forming an $\epsilon$-net over the gates. This comes at the cost of additional $\log(1/\epsilon)$ factors in the bound of \autoref{thm:fermion_NLTS}. 
\end{rmk}

\begin{rmk}\label{rem:gaussian_state}
    A similar union bound can be applied to show that any state from the set of Gaussian states cannot be a near ground state for the SYK Hamiltonian for $q\geq 4$, since an $\epsilon$-net over the set of Gaussian states has cardinality $\exp( \tilde{\mathcal{O}}(n^2 + \operatorname{poly}\log(1/\epsilon)))$. This reproduces results from prior works \cite{PhysRevResearch.3.023020,herasymenko2023optimizing}.
\end{rmk}

\subsection{Relation to NLTS results}

Our circuit lower bound is closely related to the study of  `no low-energy trivial states' (NLTS) Hamiltonians. Introduced in \cite{freedman2013quantum}, a Hamiltonian $H = \sum_i g_i \vA_i$ has the NLTS property if there is no constant-depth circuit preparing a state whose energy is above the ground energy by less than some constant fraction of the $\ell_1$ norm $\sum_i |g_i|$. Such Hamiltonians were first proven to exist in \cite{anshu2023nlts} using quantum LDPC codes.

The circuit lower bounds we give are not quite comparable to the traditional notion of NLTS. This is because we compare the energy of our low-energy states to the Hamiltonian's maximum eigenvalue rather than the $\ell_1$ norm of the coefficients. Unlike the quantum code Hamiltonian studied in \cite{anshu2023nlts,herasymenko2023fermionic}, the SYK model is highly frustrated and thus the operator norm and $\ell_1$ norms have vastly different scalings: $\Theta(n^2)$ and $\Theta(\sqrt{n})$, respectively.

Despite these differences from the standard NLTS setting, the circuit lower bounds we can establish are much stronger in two ways when compared to current progress on NLTS~\cite{8104078,10.1063/5.0113731,anschuetz2024combinatorialnltsoverlapgap,herasymenko2023fermionic,anshu2023nlts}. First, our circuit lower bounds hold for states at \emph{any} energy which is a constant fraction of the ground state energy, rather than for states below some constant-fraction energy threshold. Second, we can achieve arbitrary polynomial circuit depth lower bounds, whereas current constructions of NLTS only give a logarithmic depth lower bound.

\subsection{Other notions of non-triviality}

Though our focus so far has been on quantum circuit lower bounds, our results readily generalize to lower bounds for other classes of ansatzes via the construction of covering nets.
We begin by discussing tensor networks, focusing on matrix product states (MPSs) as a particular example. Implemented at any finite precision the number of configurations of a matrix product state on $n$ sites grows with the \emph{bond dimension} $\chi$ as $\left\lvert\left\{\ket{\psi_j}\right\}\right\rvert = \exp\left(\operatorname{\Theta}\left(\chi^2+\log n\right)\right)$.
It is thus apparent from the same argument that the minimum bond dimension such that there is an MPS achieving an energy $t\sqrt{n}$ is
\begin{equation}
    \chi=\operatorname{\Omega}_q\left(n^{q/4+1/2}t\right)
\end{equation}
with high probability.
Similarly, a classical neural network representation of the state with $W$ weights has a number of configurations growing as $\left\lvert\left\{\ket{\psi_j}\right\}\right\rvert=\exp\left(\operatorname{\Theta}\left(W\right)\right)$,
yielding the growth condition to achieve an energy $t\sqrt{n}$ with high probability:
\begin{equation}
    W=\operatorname{\Omega}_q\left(n^{q/2+1}t^2\right).
\end{equation}

\subsection{Product state approximations for spin Hamiltonians} \label{sec:product}

It is worth pointing out that there cannot be a $k$-local spin Hamiltonian with the property in \autoref{thm:fermion_NLTS}. For any traceless $k$-local spin Hamiltonian $\vH$, there is a product state achieving energy at least $\lambda_{\max}(\vH) / 3^k$. The argument is imported from \cite[proof of Theorem 2]{bravyi2019approximation}, and the proof technique bears a remarkable resemblence to the classical shadows protocol \cite{huang2020predicting}, which provides a learning algorithm for $k$-local spin operators.

\begin{prop} \label{prop:spin_product_approx}
For any $k$-local Hamiltonian $\vH$ on $n$ qubits, there is a product state $|\psi\rangle$ achieving energy
\begin{equation}
\langle\psi|\vH|\psi\rangle \geq \lambda_{\max}(\vH) / 3^k.
\end{equation}
\end{prop}
\begin{proof}
Let $|\phi\rangle$ be the true (possibly entangled) maximum-energy state achieving
\begin{equation}
\langle\phi|\vH|\phi\rangle = \lambda_{\max}
\end{equation}
For each qubit, pick a random basis out of $\{\sigma_X,\sigma_Y,\sigma_Z\}$, and measure in this basis. This gives a product state of single-qubit stabilizer states. Let $\rho$ be the resulting ensemble of pure product states we will analyze $\Tr\left(\vH\rho\right)$. It turns out that
\begin{equation}
\rho = \mathcal{E}_{1/3}^{\otimes n} (|\phi\rangle\langle\phi|)
\end{equation}
where $\mathcal{E}_p$ is the depolarizing channel
\begin{equation}
\mathcal{E}_p(\tau) = p \tau + (1-p) \mathbbm{1}
\end{equation}
Thus
\begin{equation}
\Tr\left(\vH \rho\right) = \Tr\left(\vH \mathcal{E}_{1/3}^{\otimes n}(|\phi\rangle\langle\phi|)\right) = \Tr\left(\mathcal{E}_{1/3}^{\otimes n}(\vH) |\phi\rangle\langle\phi|\right) = \langle\phi|\big(\vH / 3^k\big)|\phi\rangle = \lambda_{\max}(\vH) / 3^k
\end{equation}
Here we used that $\vH$ is traceless and $k$-local.
\end{proof}

\end{document}